\documentclass[useAMS,usenatbib,usegraphicx]{mn2e}
\usepackage{color,comment,amsmath,amsbsy,amssymb}

\usepackage{fixltx2e} 

\newcommand{\Msun}{\,\rm M_{\sun}}

\newcommand{\f}{\frac}

\newcommand{\fbreak}{f_{\rm LyC}}
\newcommand{\Lnuunits}{{\rm erg \, s^{-1} \, Hz^{-1}}}

\newcommand{\Mlim}{M_{\rm lim}}
\newcommand{\QHII}{Q_{\rm HII}}
\newcommand{\fesc}{f_{\rm esc}}
\newcommand{\taue}{\tau_{\rm e}}

\newcommand{\zetaion}{\zeta_{\rm ion}}
\newcommand{\Lya}{Ly$\alpha$}
\newcommand{\ndotion}{\dot{n}_{\rm ion}}

\begin{document}

\title[Concordance models of reionization]{Concordance models of reionization: implications for faint galaxies and escape fraction evolution}

\author[Kuhlen \& Faucher-Gigu\`{e}re]
{Michael~Kuhlen\thanks{mqk@astro.berkeley.edu},
Claude-Andr\'{e} Faucher-Gigu\`{e}re\thanks{Miller Fellow; cgiguere@berkeley.edu} \\
Theoretical Astrophysics Center, University of California, Berkeley, CA 94720} 

\maketitle

\begin{abstract}
Recent observations have constrained the galaxy ultra-violet (UV) luminosity function up to $z \sim 10$. 
However, these observations alone allow for a wide range of reionization scenarios due to uncertainties in the abundance of faint galaxies and the escape fraction of ionizing photons. We show that requiring continuity with post-reionization ($z<6$) measurements, where the \Lya\ forest provides a complete probe of the cosmological emissivity of ionizing photons, significantly reduces the permitted parameter space. 
Models that are simultaneously consistent with the measured UV luminosity function, the Thomson optical depth to the microwave background, and the \Lya\ forest data require either: 1) extrapolation of the galaxy luminosity function down to very faint UV magnitudes $M_{\rm lim}\sim-10$, corresponding roughly to the UV background suppression scale; 2) an increase of the escape fraction by a factor $\gtrsim10$ from $z=4$ (where the best fit is 4\%) to $z=9$; or 3) more likely, a hybrid solution in which undetected galaxies contribute significantly and the escape fraction increases more modestly. 
Models in which star formation is strongly suppressed in low-mass, reionization-epoch haloes of mass up to $M_{\rm h}\sim 10^{10}$ M$_{\odot}$ (e.g., owing to a metallicity dependence) are only allowed for extreme assumptions for the redshift evolution of the escape fraction. 
However, variants of such models in which the suppression mass is reduced (e.g., assuming an earlier or higher metallicity floor) are in better agreement with the data. 
Interestingly, concordance scenarios satisfying the available data predict a consistent redshift of 50\% ionized fraction $z_{\rm reion}(50\%)\sim 10$. 
On the other hand, the \textit{duration} of reionization is sensitive to the relative contribution of bright versus faint galaxies, with scenarios dominated by faint galaxies predicting a more extended reionization event. 
Scenarios relying too heavily on high-redshift dwarfs are disfavored by kinetic Sunyaev-Zeldovich measurements, which prefer a short reionization history. 
\end{abstract}

\section{Introduction}
The installation of the \textit{Wide Field Camera 3} (WFC-3) on the \textit{Hubble Space Telescope} (HST) has recently improved the efficiency of searches for faint $z\gtrsim7$ galaxies by more than an order of magnitude \citep[e.g.,][]{bouwens_discovery_2010, mclure_galaxies_2010, bunker_contribution_2010}. 
As a result, deep WFC-3 observations have provided new measurements of the rest frame ultra-violet (UV, $\sim$1,500~\AA) galaxy luminosity function at these redshifts. 
These measurements are particularly important since galaxies are the most likely sources of hydrogen reionization (e.g., Madau et al. 1999; Faucher-Gig\`ere et al. 2008a,b)\nocite{madau_radiative_1999, faucher-giguere_flat_2008, faucher-giguere_evolution_2008}. 
Nevertheless, it is difficult to robustly translate these measurements into predictions of the reionization history, because of significant uncertainties in the spectral energy distribution (SED) of the galaxies, the fraction of ionizing photons that escape into the intergalactic medium (IGM), and in the contribution of fainter, as of yet undetected galaxies. 

Because of these uncertainties, it has been unclear whether star-forming galaxies can actually reionize the Universe by $z\sim 6$ (as required by the transmission of the Ly$\alpha$ forest at lower redshifts; Fan et al. 2002; Becker et al. 2007\nocite{fan_evolution_2002, becker_evolution_2007}, although see McGreer et al. 2011\nocite{mcgreer_first_2011}) and account for the Thomson scattering optical depth to the microwave background implied by the latest, 7-year \textit{Wilkinson Microwave Anisotropy Probe} analysis \citep[WMAP-7;][]{komatsu_seven-year_2011}, corresponding to a redshift of instantaneous reionization $z_{\rm reion}=10.4\pm1.2$.\footnote{In reality, the epoch of reionization is expected to be extended in time and the Thomson scattering optical depth only provides an integral constraint on reionization.} 
Even if galaxies are in fact the dominant re-ionizing sources, it is not clear to what extent faint sources below the detection limit of existing observations are needed. 

The amount of star formation taking place in low-mass dark matter haloes is not only relevant for reionization, but also for our understanding of galaxy formation and evolution in general. 
Indeed, several lines of evidence suggest that star formation in such haloes is suppressed, at least in certain regimes. 
For instance, it is well known at lower redshifts that the baryonic mass fraction in low-mass haloes is strongly suppressed relative to $\Omega_{\rm b}/\Omega_{\rm m}$ \citep[e.g.,][]{conroy_connecting_2009, guo_how_2010}. 
This baryon deficiency is commonly attributed to a combination of feedback processes, such as galactic winds, and suppression by the photo-ionizing background \citep[e.g.,][]{dekel_origin_1986, efstathiou_suppressing_1992, murray_maximum_2005, faucher-giguere_baryonic_2011}. 
Observationally, there also appear to be far fewer dwarf galaxies in the haloes of the Milky Way and M31 than the number of dark matter sub-haloes capable of hosting them predicted in $N-$body simulations \citep[e.g.,][]{bullock_reionization_2000, madau_fossil_2008}, suggesting that some process inhibited star formation in the dark sub-haloes. 
Recently, theoretical models have also suggested that star formation may be specifically suppressed in low-mass haloes at early times due to a metallicity dependence of the star formation efficiency \citep[][]{robertson_molecular_2008, gnedin_kennicutt-schmidt_2010, krumholz_metallicity-dependent_2011, kuhlen_dwarf_2012}. 
If star formation is indeed strongly suppressed in early dwarf galaxies, then it may not be possible to rely on them to reionize the Universe. It is thus necessary to clarify the importance of those galaxies for reionization. 

The primary goal of this paper is to examine the existing observational constraints on hydrogen reionization and its sources, and to systematically determine which scenarios are (and are not) allowed by the data. 
A main distinction of our study relative to recent analyses \citep[e.g.,][]{bouwens_lower-luminosity_2011, shull_critical_2011, jaacks_steep_2012} is the inclusion of lower-redshift Ly$\alpha$ forest data \citep[see also][]{miralda-escude_evolution_2003, bolton_observed_2007, faucher-giguere_evolution_2008, pritchard_constraining_2010, haardt_radiative_2011}. 
The mean transmission of the Ly$\alpha$ forest, which is set by a balance between the ionizing background and recombinations, has the advantage of being a complete probe of the ionizing sources. 
The total instantaneous rate of injection of ionizing photons into the IGM can be measured from the Ly$\alpha$ forest without recourse to assumptions on the escape fraction or extrapolating the contribution of faint sources, two of the principal uncertainties affecting traditional analyses based on the galaxy UV luminosity function. 
We also include recent constraints on the duration of reionization from measurements of the kinetic Sunyaev-Zeldovich (kSZ) effect by the \emph{South Pole Telescope}\footnote{http://pole.uchicago.edu} (SPT) high-resolution microwave background experiment (Zahn et al. 2011; for a recent parameter space study of the kSZ signal from patchy reionization, see Mesinger et al. 2011\nocite{zahn_comic_2011, mesinger_kinetic_2011}). 

While the Ly$\alpha$ forest data are mostly restricted to $z \leq 6$, when reionization is probably complete,\footnote{Because reionization is predicted to be highly inhomogeneous, existing constraints have not ruled out that some regions of the Universe may have been reionized as late as $z\sim5$ \citep[e.g.,][]{mcgreer_first_2011}, but as we show in this paper various data taken collectively suggest that the bulk of reionization occured significantly earlier.} they provide valuable constraints in two ways. 
First, realistic reionization scenarios should continuously connect to the post-reionization IGM probed by the forest. 
Second, measurements of the galaxy UV luminosity function (analogous to those directly probing the epoch of reionization) are available over the full redshift interval covered by the Ly$\alpha$ forest data \citep[e.g.,][]{bouwens_uv_2007, reddy_steep_2009}. 
Where the data overlap, comparison of the Ly$\alpha$ forest and the UV luminosity function allows us to constrain the escape fraction and limiting magnitude (minimum luminosity) down to which the luminosity function must be integrated in order to account for all the ionizing photons measured using the forest (Faucher-Gigu\`ere et al. 2008a)\nocite{faucher-giguere_evolution_2008}. 
Since these parameters are constrained where the data overlap, we can test whether they must evolve with redshift in order to accommodate the reionization constraints from WMAP and galaxy surveys. 
Such evolution, in particular in the escape fraction, is sometimes invoked to support the hypothesis that galaxies can indeed reionize the Universe \citep[e.g.,][]{haardt_radiative_2011}, but there is little direct evidence for the required change because direct measurements of escaping Lyman continuum photons are prohobitive during the epoch of reionization. 

The plan of this paper is as follows. 
In \S \ref{sec:UF_LV}, we review how UV luminosity function measurements can be converted into predictions for the reionization history. 
We show how uncertainties in the SED of galaxies, their escape fraction, and the limiting magnitude introduce large degeneracies and allow a wide range of scenarios to be consistent with the standard WMAP constraint. 
In \S \ref{sec:LyaF}, we introduce the Ly$\alpha$ forest constraints on the ionizing background at $2 \leq z \leq 6$ and explain how these constraints relate to the ionizing sources. 
In \S \ref{sec:evolution}, we compare with the galaxy UV luminosity function and Ly$\alpha$ forest data to constrain the escape fraction and limiting magnitude at $z=4$. 
We then combine these constraints with the higher-redshift galaxy survey data and the measured WMAP optical depth to quantify the allowed scenarios, parameterized by the required limiting magnitude and evolution of the escape fraction. 
We conclude with a discussion of the implications for galaxy formation and experiments aimed at probing the epoch of reionization in \S \ref{sec:conclusions}.

Throughout, we assume cosmological parameters consistent with the WMAP 7-year data in combination with supernovae and baryonic acoustic oscillations: $(\Omega_{\rm m},~\Omega_{\rm b},~\Omega_{\Lambda},~h)=(0.28,~0.046,~0.72,~0.7)$ \citep[abbreviated \mbox{WMAP-7};][]{komatsu_seven-year_2011}. 
We adopt hydrogen and helium mass fractions $X=0.75$ and $Y=0.25$, respectively. 
All magnitudes are in the AB system \citep[][]{oke_secondary_1983}. 
Unless otherwise noted, all errors are $1\sigma$. 

\begin{figure}
\includegraphics[width=\columnwidth]{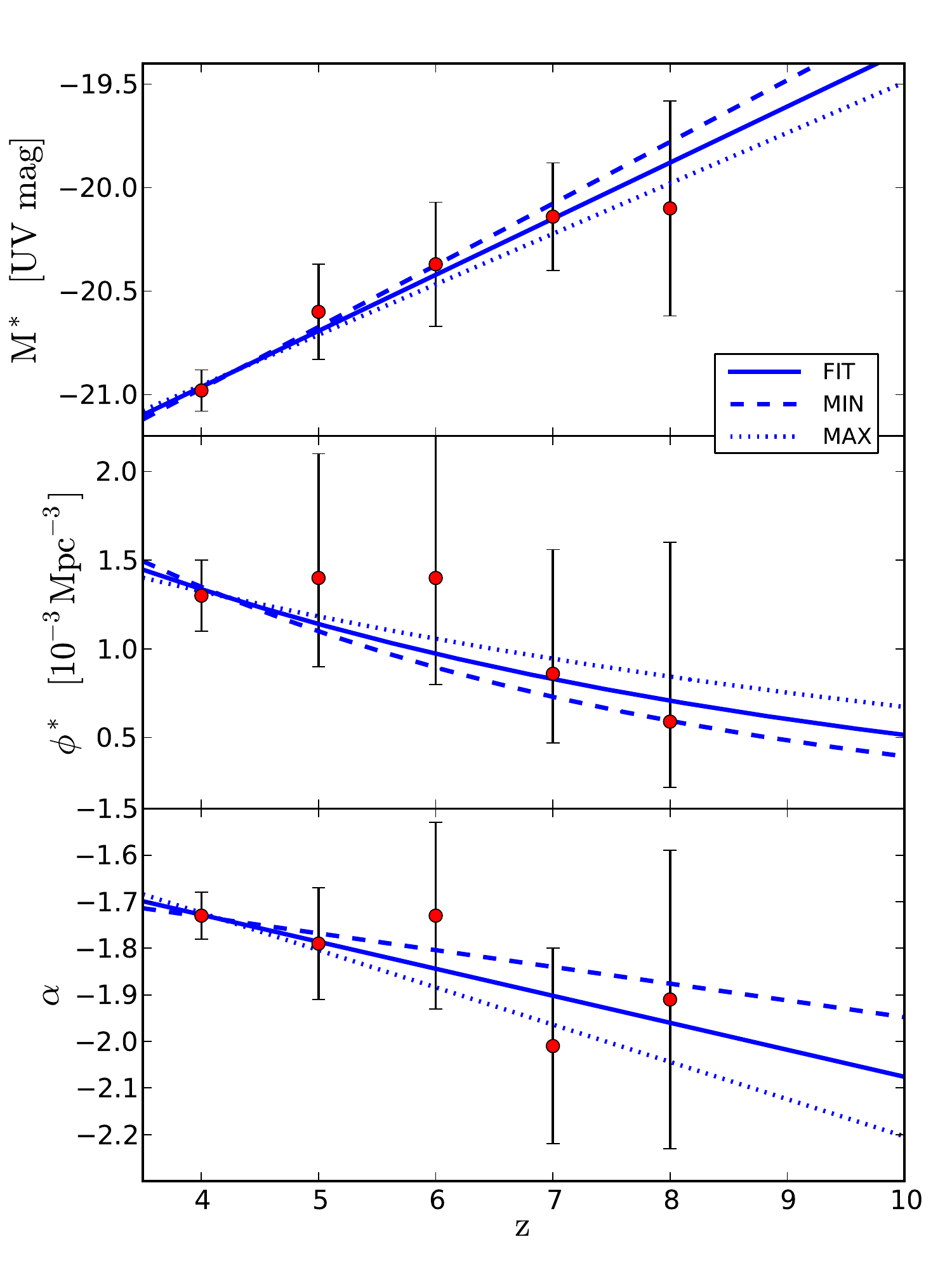}
\caption{Fits to Schechter galaxy UV luminosity function parameters versus redshift. Data points with error bars are from \citet{bouwens_lower-luminosity_2011}. The solid line is our best fit linear model (FIT). The dashed and dotted lines show the MIN and MAX models, in which the parameters were adjusted within the linear fit formal 1$\sigma$ errors to minimize (MIN) or maximize (MAX) the contribution from faint galaxies (see text for details and Table \ref{tab:LF_models} for numerical values). }
\label{fig:LF_models}
\vspace*{0.1in}
\end{figure}

\section{Galaxy survey and WMAP reionization constraints}
\label{sec:UF_LV}
The two most basic observational constraints on hydrogen reionization are the high-redshift galaxy UV luminosity function (LF) and the Thomson (electron) scattering optical depth to the microwave background measured by \mbox{WMAP-7}, $\taue=0.088 \pm 0.015$ \citep{komatsu_seven-year_2011}. 
In the following, we review how these measurements can be combined to constrain parameters of the ionizing source population, in particular the limiting UV magnitude $M_{\rm lim}$ and the escape fraction of ionizing photons $f_{\rm esc}$ from star-forming galaxies. 
The procedure is based on calculating, for a given set of assumptions on the galaxy population, the predicted evolution of the IGM ionized fraction versus redshift and evaluating the corresponding $\tau_{\rm e}$. 
As we show in \S \ref{sec:zeta_ion}, the escape fraction is degenerate with the ratio of $1,500$~\AA~UV continuum to ionizing flux, a quantity sensitive to the SED of the galaxies and whose effect we encapsulate in a dimensionless parameter $\zeta_{\rm ion}$ defined below. 
Thus, our analysis formally constrains the combination $\zeta_{\rm ion} f_{\rm esc}$. 
For simplicity, though, we will occasionally summarize our results in terms of $f_{\rm esc}$ (for values of $\zeta_{\rm ion}$ motivated by stellar population synthesis models), since it is the most uncertain of the two factors.

In this paper, we assume that the majority of the ionizing photons are produced by star-forming galaxies dominated by ordinary Pop II stars. 
In principle, other sources such as massive Pop III stars \citep[e.g.,][]{bromm_forming_1999, yoshida_era_2004}, accreting black holes \citep[e.g.,][]{haiman_observational_1998, madau_early_2004, kuhlen_first_2005}, or annihilating dark matter \citep[e.g.,][]{belikov_how_2009} could also contribute ionizing photons. 
However, there is essentially no observational support for these more exotic scenarios. 
In particular, the luminosity function of luminous quasars drops sharply beyond $z\sim2$ \citep{hopkins_observational_2007} and theoretical models suggest that only one or two supernovae from Pop III stars suffice to trigger the transition to Pop II in an early halo \citep{wise_birth_2010}. 
This is supported by IGM metallicity measurements at $z-5-6$, which show that the relative abundances are consistent with measurements down to $z\sim2$, and thus that there is no evidence for significant metal production from Pop III stars in the first billion years \citep[][]{becker_iron_2011}. 
In contrast, star-forming galaxies are now routinely observed at $z\gtrsim7$ and we show explicitly in this work that scenarios in which they are solely responsible for hydrogen reionization are consistent with the available data \citep[for a recent review, see also][]{robertson_early_2010}. 

If sources other than star-forming galaxies dominated hydrogen reionization, then the constraints on $M_{\rm lim}$ and $f_{\rm esc}$ that follow would be arbitrarily weakened. However, following Occam's razor, we do not consider such scenarios further here. 

\subsection{Calculation of the HII volume filling fraction and of the Thomson optical depth}
The evolution of the volume filling fraction of ionized hydrogen, $\QHII(z)$, is given by the differential equation
\begin{equation}
\frac{d\QHII}{dt} = \frac{\ndotion}{\bar{n}_{\rm H}} - \frac{\QHII}{\bar{t}_{\rm rec}},
\label{eq:QHII}
\end{equation}
consisting of a source term proportional to the ionizing emissivity and a sink term due to recombinations \citep{madau_star_1998}. 

Under the assumption that galaxies provide the bulk of the ionizing photons, the comoving ionizing emissivity (in units of photons per unit time, per unit volume) can be expressed as an integral over the galaxy UV LF, $\phi(M_{\rm UV})$:
\begin{equation}
\ndotion^{\rm com} = \int_{\Mlim}^\infty \!\!\! dM_{\rm UV} \, \phi(M_{\rm UV}) \gamma_{\rm ion}(M_{\rm UV}) \, \fesc \, .
\label{eq:ndotion}
\end{equation}
We denote by $\gamma_{\rm ion}(M_{\rm UV})$ the ionizing luminosity (in units of photons per unit time) of a galaxy with absolute rest-frame UV (1500\,\AA) magnitude $M_{\rm UV}$.
$\fesc$ denotes the effective escape fraction, which by definition we treat as a function of $z$ only (see \S \ref{sec:fesc}). 
The volume averaged recombination time is given by 
\begin{eqnarray}
\bar{t}_{\rm rec} & = & \f{1}{C_{\rm HII} \alpha_{\rm B}(T_{0}) \, \bar{n}_{\rm H}  (1+Y/4X) \, (1+z)^3 \, } \\
                 & \approx & 0.93 \; {\rm Gyr} \, \left( \f{C_{\rm HII}}{3} \right)^{-1} \left(\f{T_{0}}{2 \times 10^4 \, {\rm K}}\right)^{0.7} \!\! \left( \f{1+z}{7} \right)^{-3}, \nonumber
\label{eq:trec}
\end{eqnarray}
where $\alpha_B$
is the case B hydrogen recombination coefficient, $T_{0}$ is the IGM temperature at mean density, $C_{\rm HII}$ is the effective clumping factor in ionized gas, and $\bar{n}_{\rm H}$ is the mean comoving hydrogen number density.
We assume that helium is singly ionized at the same time as hydrogen, but only fully ionized later through the action of quasars (e.g., Faucher-Gigu\`ere et al. 2008a\nocite{faucher-giguere_evolution_2008}). We use the effective clumping factor to account for both the actual clumpiness of the gas and for the fact that the IGM temperature (and hence the proper recombination coefficient) in general depends on density, so that formally an average over the temperature distribution should be performed. 

The clumping factor must be selected with care, since formal averages $C_{\rm HII} \sim \langle n_{\rm HII}^{2} \rangle / \langle n_{\rm HII} \rangle^{2}$ over simulation volumes yield large values $\sim30$ \citep[e.g.,][]{gnedin_reionization_1997, springel_history_2003} that imply very demanding requirements on the ionizing sources. 
These large clumping factors arise because the average includes very dense galaxy halo gas. 
However, absorption of ionizing photons by gas inside (or in the immediate vicinity of) galaxies is already accounted for by the escape fraction. 
Thus, the correct clumping factor to use is one that accounts only for recombinations occurring in the more diffuse IGM. 
Although some ambiguity is inherent in this definition, recent studies suggest that values $C_{\rm HII}=1-3$ are appropriate during the epoch of reionization \citep[e.g.,][]{pawlik_keeping_2009, shull_critical_2011, mcquinn_lyman_2011}. 
The IGM temperature $T_{0}$ is also uncertain, but the fiducial value $T_{0}=2\times10^{4}$ K is reasonable for freshly reionized gas \citep[][]{hui_thermal_2003}. 

The Thomson optical depth to microwave background is then obtained by integrating $\QHII$,
\begin{equation}
\taue = \int_0^\infty dz \frac{c (1+z)^2}{H(z)} \QHII(z) \, \sigma_{\rm T} \, \bar{n}_{\rm H} \, (1 + \eta Y/4X),
\label{eq:taue}
\end{equation}
where $H(z)$ is the Hubble parameter, $\sigma_{\rm T}$ is the Thomson cross section, and we consider helium to be only singly ionized ($\eta=1$) at $z>4$ and doubly ionized ($\eta=2$) at lower redshift.

The main uncertainties in these calculations, which we discuss next, are
\begin{enumerate}
\item the extrapolation of the LF to magnitudes and redshifts for which no direct measurement exists (\S~\ref{sec:highz_LF}),
\item the conversion from $M_{\rm UV}$ to ionizing photon luminosity ($\gamma_{\rm ion}$; \S~\ref{sec:zeta_ion}),
\item and the escape fraction of ionizing photons ($f_{\rm esc}$; \S~\ref{sec:fesc}). 
\end{enumerate}

\subsection{High redshift galaxy luminosity functions}
\label{sec:highz_LF}

We base our analysis on recent observational determinations of the rest-frame UV LF at $z\geq4$ in the HUDF09 \citep{beckwith_hubble_2006,oesch_udf05_2007}, ERS \citep{windhorst_hubble_2011}, and CANDELS fields \citep{grogin_candels:_2011,koekemoer_candels:_2011} by \citet{bouwens_uv_2007,bouwens_ultraviolet_2011,bouwens_lower-luminosity_2011}. The best-fit Schechter function parameters ($\phi^*$, $M^*$, and $\alpha$) are summarized in \citet[][hereafter B11]{bouwens_lower-luminosity_2011}. $M^{*}$ quantifies the characteristic magnitude, $\phi^{*}$ measures the comoving number density, and $\alpha$ is the faint-end slope. 

\begin{figure}
\includegraphics[width=\columnwidth]{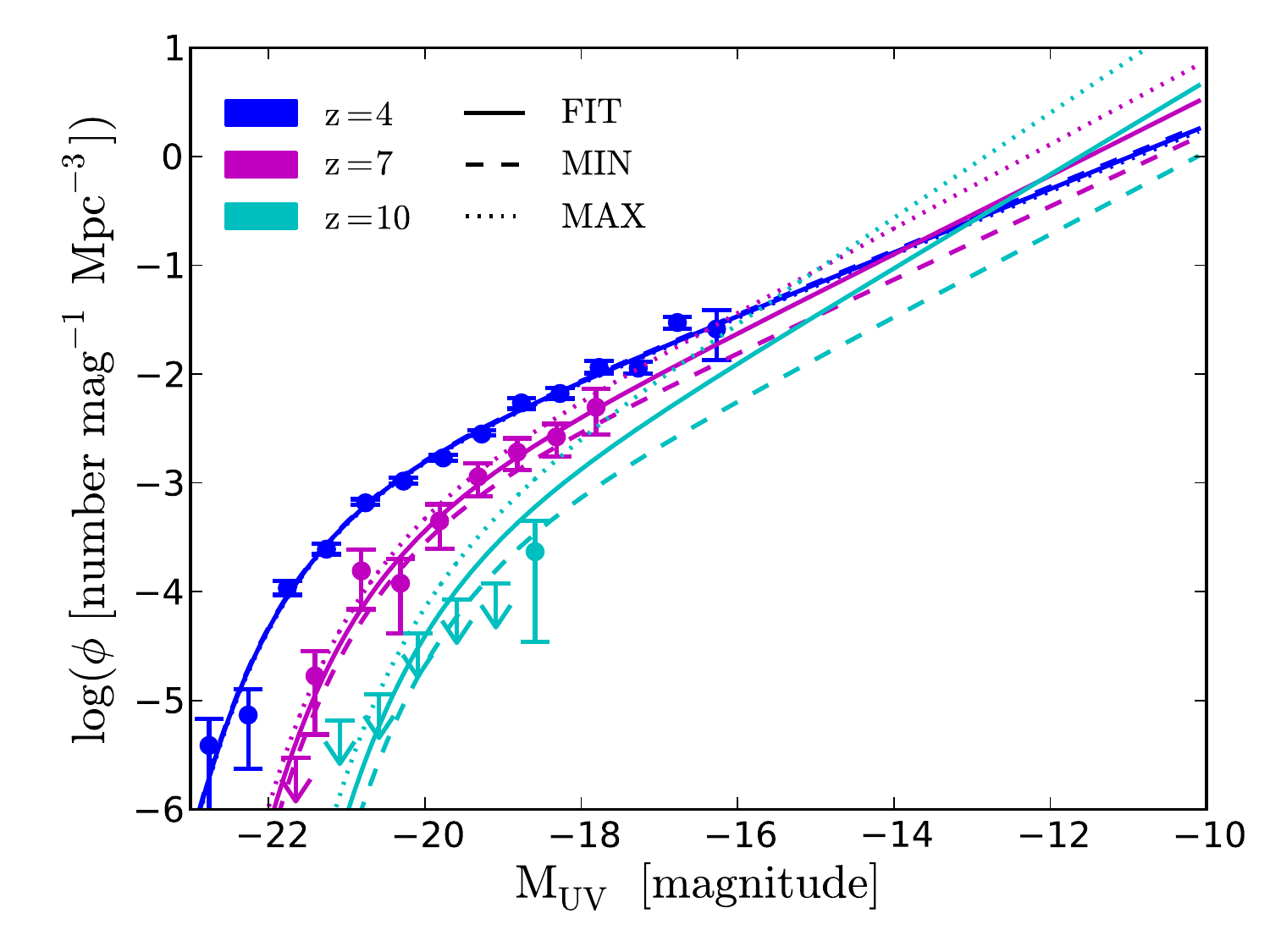}
\caption{Comparison of our FIT, MIN, and MAX luminosity function models to the data of \citet{bouwens_candidate_2011} at $z=4$ and 7, and to the updated limits from \citet{oesch_expanded_2011} at $z=10$. The $z=10$ data points, obtained from a single galaxy candidate, were not included in the fits. 
}
\label{fig:LF_z=4,7,10}
\vspace*{0.1in}
\end{figure}

\begin{table*}
\begin{minipage}{\textwidth}
\begin{center}
 \caption{UV luminosity function evolution models}
 \label{tab:LF_models}
 \begin{tabular}{lcccccl}
  \hline
  Model & \multicolumn{2}{c}{$M^*$} & \multicolumn{2}{c}{$\log_{10} \phi^*$} & \multicolumn{2}{c}{$\alpha$}  \\
       & $A$ & $B$ & $A$ & $B$ & $A$ & $B$ \\
  \hline
  FIT & $-20.42 \pm 0.05$ & $0.27 \pm 0.03$ & $-3.01 \pm 0.04$ & $-0.07 \pm 0.02$ & $-1.84 \pm 0.04$ & $-0.06 \pm 0.02 $ \\ 
  MIN & $-20.37         $ & $0.30         $ & $-3.05         $ & $-0.09         $ & $-1.80          $ & $-0.04 $ \\ 
    MAX & $-20.47         $ & $0.24         $ & $-2.97         $ & $-0.05         $ & $-1.88          $ & $-0.08 $ \\ 
      \hline
 \end{tabular}

 \medskip
 FIT denotes the best linear fit of the form $\{ M^{*},~\log_{10}{\phi^{*}},~\alpha\}= A+B(z-6)$ to the Schechter parameters reported in \cite{bouwens_lower-luminosity_2011} at $z=4,~5,~6,~7$ and 8. The parameters of the MAX and MIN models are adjusted within 1$\sigma$~of the best fit (independently) so as to maximize and minimize the contribution of faint galaxies. 
 \end{center}
\end{minipage}
\end{table*}

To interpolate between redshift bins and extrapolate to redshifts not directly probed by the data, we fit the redshift evolution of the three Schechter parameters to a simple linear model of the form $\{ M^{*},~\log_{10}{\phi^{*}},~\alpha\}= A+B(z-6)$. The best-fit parameters, denoted FIT, are given in Table \ref{tab:LF_models}. In order to explore the uncertainties in the extrapolation to very faint galaxies, we adopt two additional models, in which we vary the redshift evolution within the formal 1$\sigma$ errors of our linear fits to either maximize (MAX) or minimize (MIN) the contribution from faint galaxies. Compared to our FIT model, the MAX model has a slightly brighter and less rapidly dimming $M^*$, a slightly larger and more slowly decreasing $\phi^*$, and a steeper and more quickly steepening faint end slope $\alpha$; and vice-versa for the MIN model. Figure~\ref{fig:LF_models} shows our three fits for the redshift evolution of the Schechter parameters. 

Figure \ref{fig:LF_z=4,7,10} shows how these fits compare to the actual LF data from \citet{bouwens_candidate_2011} at $z=4$ and $z=7$, and to the 1$\sigma$ upper limits at $z\sim10$ obtained from the detection of a single galaxy candidate by \citet{oesch_expanded_2011}. While the FIT and MIN models are in good agreement with the $z\sim10$ limits, the MAX model predicts more galaxies than observed at $M_{\rm UV}=-19.6$ by a $\sim 2.5\sigma$. Given the substantial uncertainties in estimating limits from a single candidate in a relatively small field, we however consider the MAX model to represent a valid limiting case.

\begin{figure}
\includegraphics[width=\columnwidth]{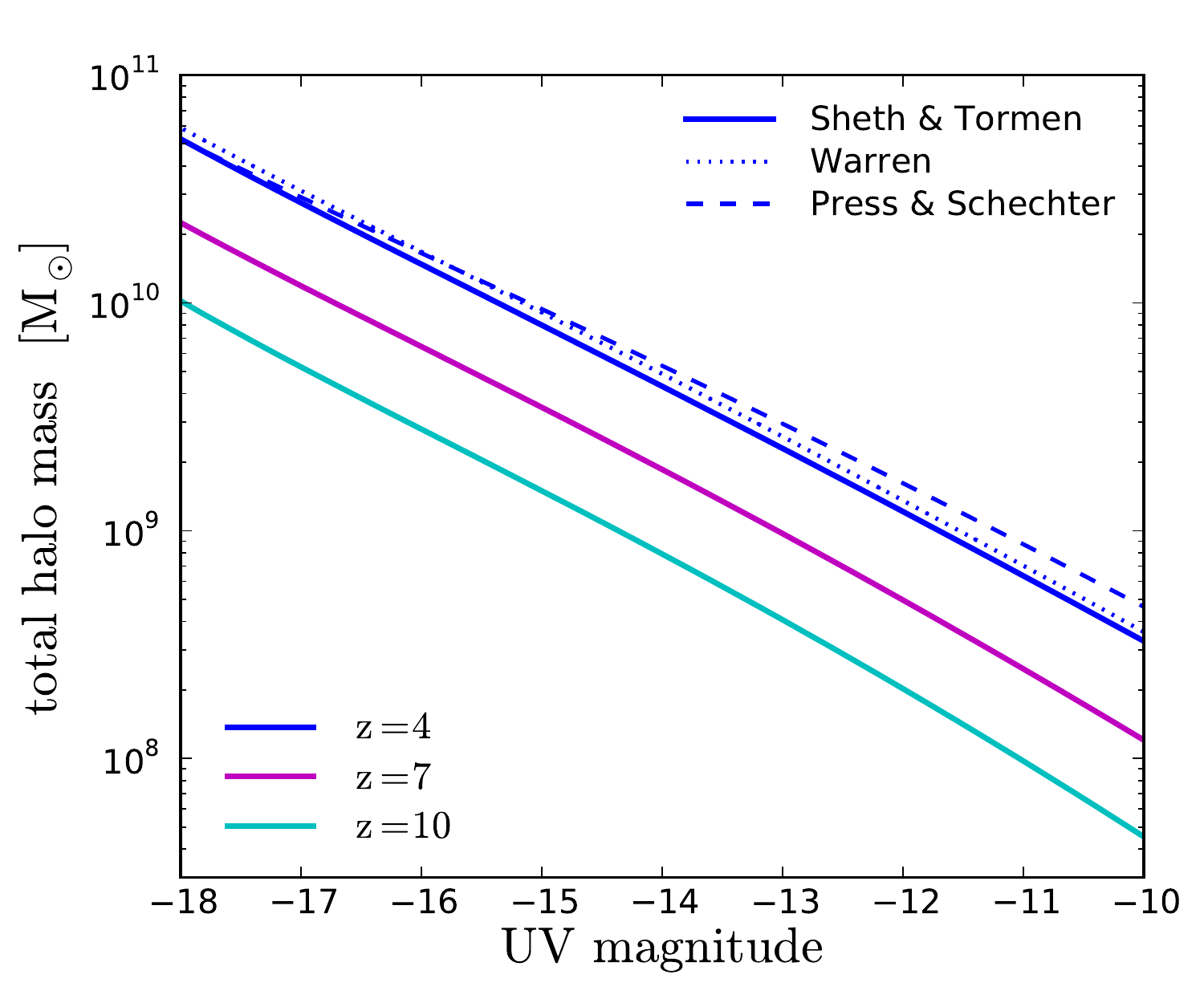}
\caption{Abundance matching between the dark matter halo mass function and the UV luminosity functions from Bouwens et al. (2010a) at $z=4,~7$ and $10$.}
\label{fig:abundance_match}
\vspace*{0.1in}
\end{figure}

For comparison with theoretical predictions, it is useful to relate the UV magnitudes to the total mass of the haloes likely to host these galaxies. Since direct mass determinations from gravitational lensing or clustering are not available at very high redshift, we attempt to establish such a relation via an abundance matching technique, by equating the cumulative dark matter halo mass function to the cumulative UV luminosity function over the redshifts of interest. At lower redshift the validity of the abundance matching technique has been demonstrated by its ability to reproduce the spatial clustering of galaxies in the SDSS/LRG catalog \citep{conroy_modeling_2006,moster_constraints_2010,guo_how_2010}. Here, we use the UV luminosity, which traces star formation rather than stellar mass, and we should expect a larger scatter in its relation to total halo mass. Nevertheless, the relation is likely to still be monotonic on average and we therefore expect the abundance matching results to be valid at the order-of-magnitude level. The results are shown in Figure.~\ref{fig:abundance_match}, which reveals that the faint values of $\Mlim$ advocated by B11 correspond to total halo masses below $10^9 \Msun$. These results are in good agreement with a similar determination by \citet{trenti_galaxy_2010}.

\subsection{Conversion from UV magnitude to ionizing luminosity}
\label{sec:zeta_ion}
To evaluate equation (\ref{eq:ndotion}), it is necessary to convert from the measured UV magnitudes to ionizing luminosity (the $\gamma_{\rm ion}$ term). 
To do so, we adopt a simple double power-law model for the galaxy SED in the relevant range,\footnote{Some authors first convert the UV magnitude to a star formation rate, then convert the star formation rate to a rate of production of ionizing photons. This however introduces extraneous steps.}
\begin{equation}
 L_\nu = L_{\nu_{1500}}
 \begin{cases}
  \left( \f{\nu}{\nu_{1500}} \right)^{\beta_\nu} &  h\nu < 1~{\rm Ry} \\ 
  \fbreak \left( \f{\nu_{912}}{\nu_{1500}} \right)^{\!\beta_\nu} \left( \f{\nu}{\nu_{912}} \right)^{-\gamma} & 1~{\rm Ry}\leq h\nu < 4\, {\rm Ry} \\ 
  0 & h\nu>4~{\rm Ry}.
 \end{cases}
 \label{eq:Lnu}
\end{equation}
This simple form is adequate to capture the main features of more detailed stellar population synthesis models over the limited energy range of interest \citep[cf.][]{leitherer_starburst99:_1999,schaerer_transition_2003}. 
We use the notation $\nu_{\lambda}$ to denote the frequency corresponding to wavelength $\lambda$/\AA, e.g. $\nu_{912}$ is the frequency at the 912~\AA~Lyman edge. 
Note that the ``$\beta$'' slopes often discussed in the literature \citep[e.g.,][]{bouwens_very_2010} are usually defined in terms of wavelength, $L_{\lambda} \propto \nu^{-\beta_{\lambda}}$, so that we have the relation $\beta_\nu = -(\beta_\lambda + 2)$. 

The hydrogen ionizing photon luminosity ($\gamma_{\rm ion}$) is then given by
\begin{equation}
\gamma_{\rm ion} = \int_{\nu_{912}}^{\infty} \! \frac{d\nu}{h \nu} L_\nu \, \equiv \, 2 \times 10^{25} \, {\rm s}^{-1} \, \left( \f{L_{\nu_{1500}}}{\Lnuunits} \right) \zetaion.
\end{equation}
To express this as a function of UV magnitude, we use the standard AB relation $\log_{10}(L_{\nu_{1500}}/(\Lnuunits)) = 0.4\,(51.63 - M_{\rm UV})$.  
Using equation~\ref{eq:Lnu}, we can solve for the dimensionless parameter $\zeta_{\rm ion}$:
\begin{equation}
\zetaion = 1.5 \left( \f{\fbreak}{0.2} \right) \, (1.65)^{\beta_\nu} \left( \f{1 - 4^{-\gamma}}{\gamma} \right).
\end{equation}
This parameter, a function of the stellar spectrum characteristics, encapsulates all the information necessary to convert from UV magnitude to ionizing photon luminosity.

In order to bracket the uncertainties in the spectral parameters $(\fbreak, \beta_\nu, \gamma)$, we consider three different models: a fiducial model (FID) with $\zetaion=1$, a harder spectrum model (HARD) with $\zetaion=2$, and a softer spectrum model (SOFT) with $\zetaion=0.5$. 
This range is representative of Pop II star-forming galaxies with continuous star formation histories and age $\sim 10-100$ Myr \citep[][]{leitherer_starburst99:_1999}. 
Note that converting from $M_{\rm UV}$ to ionizing luminosity via the star formation rate as done in B11 corresponds to $\zetaion = 1$ (our FID model). Our three $L_\nu$ models thus span a factor of two variation (up and down) around the hydrogen-ionizing luminosity used by B11.

\subsection{The escape fraction of ionizing photons}
\label{sec:fesc}
Some fraction of the ionizing radiation produced by stellar populations is absorbed by dust and neutral hydrogen within their host galaxies, and thus does not contribute to ionizing the IGM. We capture this suppression by a simple multiplicative prefactor, $\fesc$, applied in equation~(\ref{eq:ndotion}). Since our calculations are tied to the observed rest-frame UV LF, our $\fesc$ is strictly speaking a \textit{relative} escape fraction, capturing the additional suppression of photons blueward of the Lyman edge compared to 1500\,\AA\ photons. While neutral hydrogen only absorbs the ionizing photons, dust extinguishes 1500\,\AA\ and ionizing photons similarly. Because of this broad band extinction by dust, $f_{\rm esc}$ is not equal to the fraction of \textit{all} ionizing photons produced by stars which are absorbed in the galaxy. Evaluating the latter would require knowledge of dust extinction, but is not actually required for our purposes. 
Similar relative definitions of the escape fraction are often adopted observationally as well \citep[e.g.,][]{steidel_lyman-continuum_2001, shapley_direct_2006, inoue_escape_2006}.

The true escape fraction may well vary with galaxy mass, age, star formation history, or other properties. Such dependences are however essentially unknown at this time. We therefore assume in this work that $f_{\rm esc}$ is a function of $z$ only, i.e. we use $f_{\rm esc}(z)$ to represent an effective escape fraction averaged over the galaxy population at redshift $z$, suitably weighted by the (unabsorbed) ionizing luminosity. A time dependence of $\fesc$ could thus arise from either a genuine time evolution in the escape fraction of galaxies (e.g., owing to an evolution in the star formation rate and its associated feedback), or from a redshift evolution in the make up of the galaxy population, with the escape fraction of galaxies with certain properties remaining constant. In \S \ref{sec:evolution}, we quantify the redshift evolution required of $f_{\rm esc}$ required by the data for different scenarios. 

\begin{figure}
\includegraphics[width=\columnwidth]{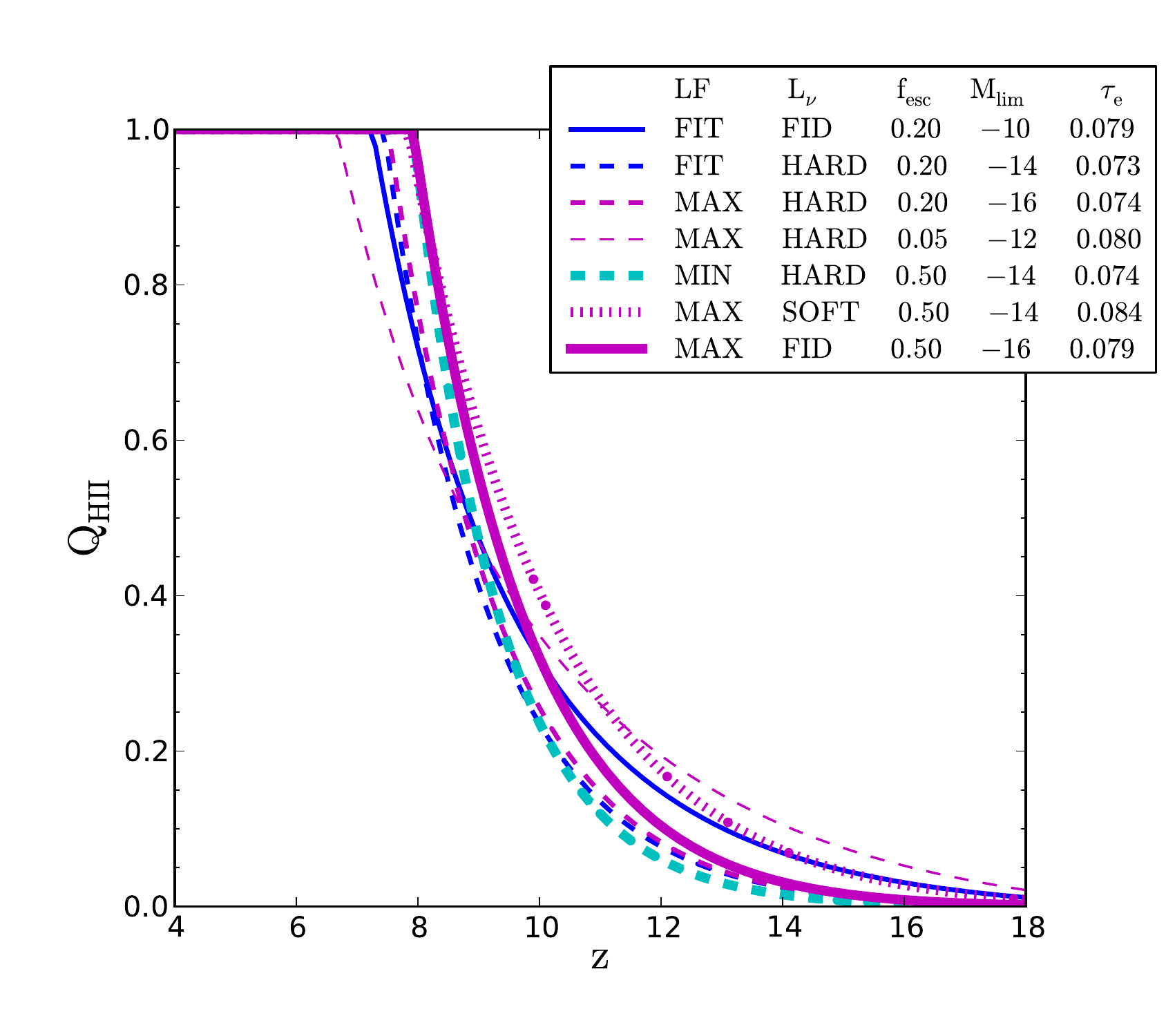}
\caption{Volume fraction filling of HII regions as a function of redshift for a set of  representative models that satisfy the measured galaxy UV LF and the \mbox{WMAP-7} Thomson scattering optical depth. LF evolution fits FIT, MIN, and MAX are shown in blue, cyan, and magenta, respectively. The FID, SOFT, and HARD spectral hardness models are indicated with solid, dotted, and dashed lines. The line thickness corresponds to $\fesc=5$\%, 20\%, and 50\% (from thin to thick).}
\label{fig:QHII_vs_z}
\vspace*{0.1in}
\end{figure}

\subsection{Range of models allowed by the UV LF and \mbox{WMAP-7} constraints alone}
\label{sec:allowed}

\begin{figure}
\includegraphics[width=\columnwidth]{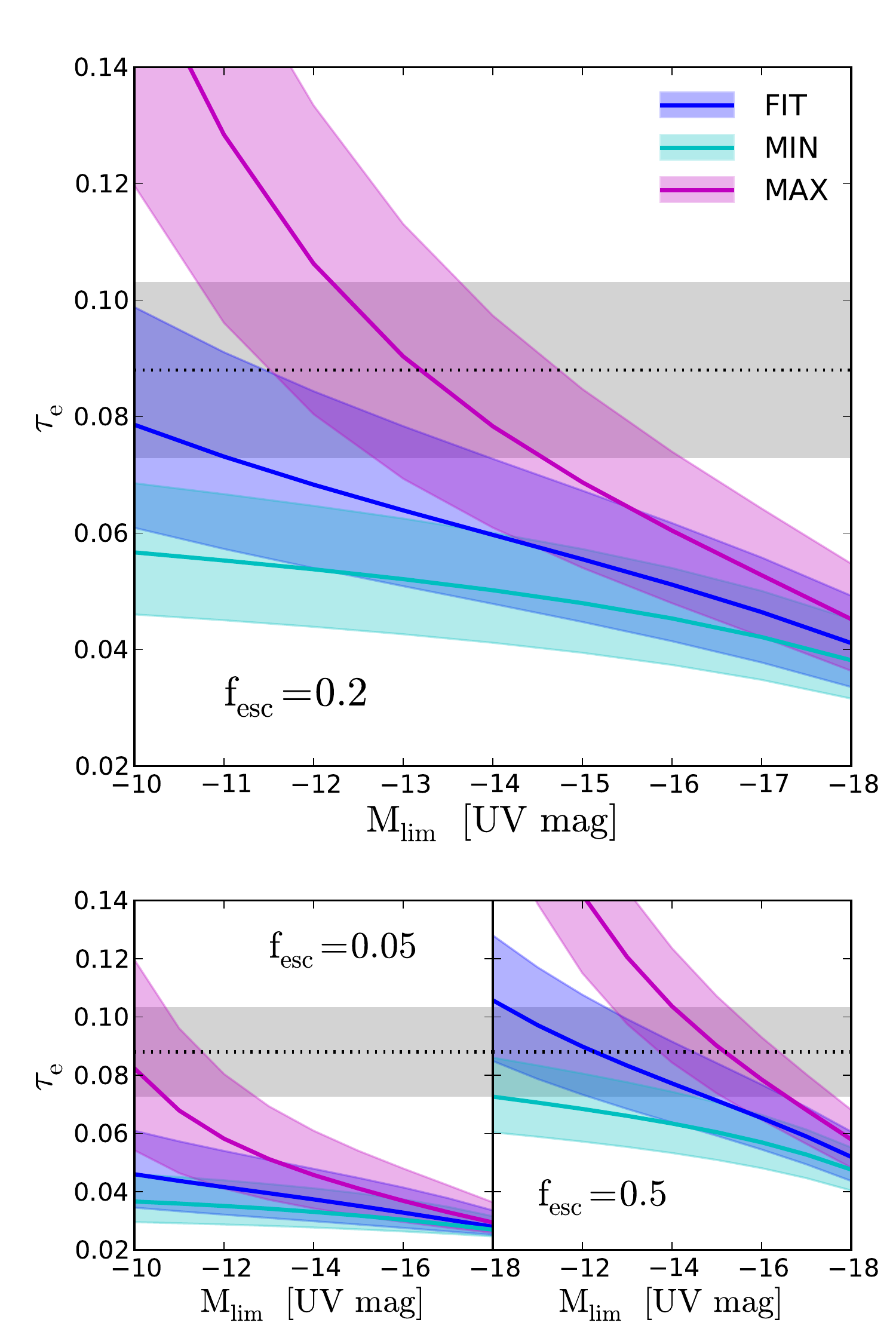}
\caption{Thomson scattering optical depth to the microwave background versus limiting UV magnitude. The colors represent our three different galaxy UV LF parameterizations: FIT (blue), MIN (cyan), and MAX (magenta). The solid line corresponds to the FID ($\zetaion=1$) $L_\nu$-model, and the shaded regions are bounded by the SOFT ($\zetaion=0.5$) and HARD ($\zetaion=2$) models. The \mbox{WMAP-7} $\taue=0.088 \pm 0.015$ \citep{komatsu_seven-year_2011} is indicated with a gray band. The top panel is for $\fesc=20\%$, the bottom left for $\fesc=5\%$ and the bottom right for $\fesc=50\%$. $\Mlim$, $\zetaion$, and $\fesc$ are assumed constant in these calculations, and we used a clumping factor of $C_{\rm HII}=3$.}
\label{fig:taue_vs_Mlim}
\vspace*{0.1in}
\end{figure}

In Figure~\ref{fig:taue_vs_Mlim} we show the Thomson optical depth, $\taue$, for different reionization scenarios consistent with the measured UV luminosity function. The models explored correspond to varying assumptions for $M_{\rm lim}$, $\zeta_{\rm ion}$, and $f_{\rm esc}$, which are further assumed here to be constant with redshift.

For the best-fit UV LF evolution parameterization (FIT), fiducial $L_\nu$-model ($\zetaion=1$, FID), and $\fesc=0.2$ (solid blue line in the top panel), we recover the result of B11 that a very faint limiting magnitude, $\Mlim \ga -11$, is required in order to produce an optical depth in agreement with \mbox{WMAP-7}. However, many other solutions are possible. For example, the same LF model with a harder spectrum (upper edge of blue shaded region) is consistent with the \mbox{WMAP-7} data for $\Mlim=-14$, and with the MAX LF model (magenta band) the \mbox{WMAP-7} $\taue$ constraint can accommodate values of $\Mlim$ ranging from $-11$ to as bright as $-16$, depending on the spectral hardness. The escape fraction provides yet another degree of freedom. With a constant $\fesc$ of 5\% (bottom left panel), most models cannot satisfy the \mbox{WMAP-7} constraint. On the other hand, if $\fesc=50\%$ even the MIN LF fit or models with very soft spectra can result in a sufficiently high $\taue$.

In Figure~\ref{fig:QHII_vs_z} we show the volume filling factor of HII regions, $\QHII(z)$, for a few representative models that all satisfy the \mbox{WMAP-7} $\taue$ constraint. Interestingly, these scenarios have limiting magnitudes ranging from $-10$ to $-16$ and include models that extrapolate the contribution of faint galaxies quite differently. We conclude that the existing measurements of the high-redshift galaxy luminosity function (still limited to relatively luminous sources) and of the Thomson optical depth to the microwave background do not uniquely determine how reionization proceeded. In particular, these constraints do not suffice to determine the role played by low-luminosity galaxies. 

\section{\Lya~forest constraints on the ionizing sources}
\label{sec:LyaF}

The Ly$\alpha$ forest provides complementary constraints on the cosmological emissivity of ionizing photons \citep[][]{miralda-escude_evolution_2003, bolton_observed_2007, faucher-giguere_evolution_2008}. Although saturation prevents accurate measurements of the Ly$\alpha$ forest at $z \ga 6$ \citep[e.g.,][]{fan_evolution_2002}, it has the advantage of being a complete probe, in the sense that it includes the contribution of all ionizing sources, even if they are individually too faint to be detected in galaxy surveys. Furthermore, the ionizing emissivity implied by the Ly$\alpha$ forest does not depend on an assumed escape fraction. Thus, the Ly$\alpha$ forest is not subject to the two main uncertainties affecting the inference of the ionizing emissivity from the galaxy UV luminosity function, namely $M_{\rm lim}$ and $f_{\rm esc}$. By assuming continuity between the post-reionization epochs probed by the Ly$\alpha$ forest and the reionization epoch probed by high-redshift galaxy surveys, it is therefore possible to significantly reduce the permitted parameter space. 
In particular, comparison of the UV luminosity function and Ly$\alpha$ data where they overlap allows us to constrain a combination of escape fraction, the limiting magnitude, and the conversion factor from 1,500~\AA~UV to ionizing luminosity (\S \ref{sec:evolution}). 

\subsection{Total ionization rate from the Ly$\alpha$ forest}
The basic quantity constrained by the \Lya~forest is the hydrogen photoionization rate $\Gamma_{\rm HI}$,
\begin{equation}
\Gamma_{\rm HI}(z) = 
4\pi \int_{\nu_{912}}^{\infty} 
\frac{d\nu}{h \nu}
 J_{\nu}(z) \sigma_{\rm HI}(\nu),
\end{equation}
where $J_{\nu}$ is the average specific intensity of the ultra-violet background, $\sigma_{\rm HI}(\nu)$ is the photoionization cross section of hydrogen, and the integral is from the Lyman limit to infinity. Indeed, the mean level of transmission of the Ly$\alpha$ forest is set by the equilibrium between the ionizing background and recombinations in the IGM. Thus, given a model of the density fluctuations in the IGM and knowledge of the intergalactic gas temperature-density relation, the mean transmission of the Ly$\alpha$ forest can be inverted to give $\Gamma_{\rm HI}$ \citep[e.g.,][]{rauch_opacity_1997}. 

In this work, we use principally the $\Gamma_{\rm HI}$ data points from Faucher-Gigu\`ere et al. (2008a,b)\nocite{faucher-giguere_flat_2008, faucher-giguere_evolution_2008} based on the mean transmission measurement of Faucher-Gigu\`ere et al. (2008d)\nocite{faucher-giguere_direct_2008}. 
This mean transmission measurement, based on 86 high-resolution and high-signal-to-noise quasar spectra covering Ly$\alpha$ redshifts $2 \leq z \leq 4.2$, was corrected for absorption by metal ions and for biases in the continuum fits, an important effect at $z\gtrsim4$. 
At $z=5$ and $z=6$, we use the constraints on $\Gamma_{\rm HI}$ from \cite{bolton_observed_2007}, also from mean transmission data. 
We do not use proximity effect measurements, as they are typically of lower statistical precision and affected by more severe systematics (e.g., Faucher-Gigu\`ere et al. 2008c\nocite{faucher-giguere_line--sight_2008}). 
Nevertheless, at $z=5-6$, where some of these effects are mitigated, the proximity effect measurements of \cite{calverley_measurements_2011} are consistent with \cite{bolton_observed_2007}.
  
In Table \ref{tab:LyaF_data}, we summarize the $\Gamma_{\rm HI}$ measurements and other inputs used in our Ly$\alpha$ forest analysis. 

\begin{table*}
 \caption{Ly$\alpha$ forest constraints on the ionizing emissivity}
 \label{tab:LyaF_data}
 \begin{tabular}{ccccc}
 \hline
  $z$ & $\Gamma_{\rm  HI}$ & $\lambda_{\rm mfp}^{912}$ & $\dot{n}_{\rm ion}^{\rm com}$ & References \\
     & $10^{-12}$ s$^{-1}$ & pMpc & $10^{50}$~s$^{-1}$ cMpc$^{-3}$ & \\
  \hline
  2.0 &  0.64$\pm$0.18 & 303$\pm$84  & 2.0$\pm$0.8 ($^{+2.1}_{-1.4}$) & FG08, SC10 \\
  2.2 &  0.51$\pm$0.10 & 227$\pm$61  & 1.7$\pm$0.6 ($^{+1.7}_{-1.2}$) & FG08, SC10  \\
  2.4 & 0.50$\pm$0.08  & 174$\pm$45  & 1.8$\pm$0.6 ($^{+1.8}_{-1.2}$)  & FG08, SC10  \\
  2.6 & 0.51$\pm$0.07  & 135$\pm$34  & 2.0$\pm$0.6 ($^{+1.9}_{-1.3}$)  & FG08, SC10  \\
  2.8 & 0.51$\pm$0.06  & 106$\pm$26  & 2.2$\pm$0.6 ($^{+2.0}_{-1.4}$)  & FG08, SC10  \\
  3.0 &  0.59$\pm$0.07 & 84.4$\pm$21  & 2.7$\pm$0.7 ($^{+2.5}_{-1.8}$)  & FG08, SC10 \\
  3.2 & 0.66$\pm$0.08  & 67.9$\pm$16  & 3.3$\pm$0.9 ($^{+3.0}_{-2.2}$)  & FG08, SC10 \\
  3.4 & 0.53$\pm$0.05  & 55.2$\pm$13  & 2.8$\pm$0.7 ($^{+2.5}_{-1.8}$)  & FG08, SC10 \\
  3.6 & 0.49$\pm$0.05  & 49.5$\pm$2.1 & 2.6$\pm$0.3 ($^{+1.7}_{-1.5}$)  & FG08, P09  \\
  3.8 & 0.51$\pm$0.04  & 41.7$\pm$2.4  & 2.8$\pm$0.3 ($^{+1.8}_{-1.6}$)  & FG08, P09  \\
  4.0 & 0.55$\pm$0.05 & 34.0$\pm$3.1 & 3.2$\pm$0.4 ($^{+2.2}_{-1.9}$)  & FG08, P09  \\
  4.2 & 0.52$\pm$0.08  & 26.2$\pm$3.9  & 3.5$\pm$0.8 ($^{+2.9}_{-2.2}$)  & FG08, P09 \\
  5.0 & 0.52$^{+0.35}_{-0.21}$  & 13.9$\pm$3.6  & 4.3$\pm$2.6 ($\pm$2.6)  & B07, SC10  \\
  6.0 & $<$0.19  &  7.0$\pm$2.0 & $<$2.6 ($<2.6$) & B07, SC10 \\
  \hline
 \end{tabular}

 \medskip
 The HI photoionization rates measurements are taken from Faucher-Gig\`ere et al. (2008a)\nocite{faucher-giguere_evolution_2008} (FG08) and \cite{bolton_observed_2007} (B07); the mean free paths are taken from the fits of \cite{prochaska_direct_2009} (P09) and \cite{songaila_evolution_2010} (SC10). 
Errors on $\Gamma_{\rm HI}$ and $\lambda_{\rm mfp}^{912}$ are $1\sigma$ and predominantly statistical (except for the B07 
$\Gamma_{\rm HI}$ points, which include a systematic error budget). 
Total uncertainties on $\dot{n}_{\rm ion}^{\rm com}$, including systematic effects arising from the spectral shape of the UV background and the thermal history of the IGM, are given in parentheses and shown by the light gray band in Figure \ref{fig:LyaF} (see the text). 
The prefixes `p' and `c' indicate proper and comoving units, respectively.
\end{table*}

\subsection{From ionization rate to ionizing emissivity}

The quantity most directly related to the sources of ionizing photons is their spatially-averaged emissivity, $\epsilon_{\nu}$ (here in proper, specific units). Assuming that the ionizing background has a power-law spectrum $J_{\nu}=J_{\nu_{912}}(\nu/\nu_{912})^{-\gamma_{\rm bg}}$ between the HI and HeII ionizing edges (and zero beyond), 
\begin{equation}
J_{\nu_{912}} =
\frac{\Gamma_{\rm HI} h (\gamma_{\rm bg}+3)}{4\pi \sigma_{\rm HI}(\nu_{912})} \left[ 1 - \frac{1}{4^{\gamma_{\rm bg}+3}} \right]^{-1}.
\end{equation}
Since
\begin{equation}
\label{local source approximation emissivity}
\epsilon_{\nu}(z)
\approx
4\pi 
\frac{J_{\nu}(z)}{\lambda_{\rm mfp}(\nu,~z)},
\end{equation}
where $\lambda_{\rm mfp}$ is the mean free path of ionizing photons in proper units (denoted $\lambda_{\rm mfp}^{912}$ at the Lyman limit),\footnote{This approximation is valid at $z\geq2$, where the mean free path is much smaller than the Hubble scale.} 
\begin{equation}
\label{epsilon LL from Gamma}
\epsilon_{\nu_{912}}(z)\approx
\frac{\Gamma_{\rm HI}(z) h (\gamma_{\rm bg}+3)}{\sigma_{\rm HI} \lambda_{\rm mfp}^{912}(z)} \left[ 1 - \frac{1}{4^{\gamma_{\rm bg}+3}} \right]^{-1}.
\end{equation}
Assuming similarly that $\epsilon_{\nu} = \epsilon_{\nu_{912}} ( \nu / \nu_{912})^{-\gamma}$ between the HI and HeII ionizing edges (and zero beyond),\footnote{Since the ionizing background spectrum is affected by filtering by the IGM, in general $\gamma \neq \gamma_{\rm bg}$ \citep[e.g.,][]{haardt_radiative_1996, faucher-giguere_new_2009}.} 
\begin{align}
\label{ndot ion}
\dot{n}_{\rm ion}^{\rm com}(z) & = \frac{1}{(1+z)^{3}} \int_{\nu_{912}}^{\infty} \frac{d\nu}{h \nu} \epsilon_{\nu}(z) \\ \notag
& = \frac{1}{(1+z)^{3}} \frac{\Gamma_{\rm HI}(z) h}{\sigma_{\rm HI} \lambda_{\rm mfp}^{912}(z)}
\frac{(\gamma_{\rm bg} + 3)}{\gamma}
\left[ 1 - \frac{1}{4^{\gamma}} \right] \\ \notag 
& ~~~~~~~~~~~~~~~~~~~~~~~~~~~~~~~~~~~~~~~~\times \left[ 1 - \frac{1}{4^{\gamma_{\rm bg}+3}} \right]^{-1}
\end{align}
(compare with eq. (\ref{eq:ndotion})). 
Equation (\ref{ndot ion}) shows how the total comoving emissivity of ionizing photons can be derived from the photoionization rate measured from the Ly$\alpha$ forest and knowledge of the mean free path of the ionizing photons, given a model for the spectral shape of the ionizing sources and their integrated background. 

At $3.6 \leq z \leq 4.2$, we use the mean free path measured by \cite{prochaska_direct_2009} using a stacking analysis. 
This approach avoids the usual uncertainties in calculating the mean free path from the column density distribution stemming from the difficulty of measuring the column density of systems near the Lyman limit (on the flat part of the curve of growth). 
At the other redshifts $2 \leq z \leq 6$, we use the mean free path derived by \cite{songaila_evolution_2010} based on a new analysis of the column density distribution. 
These expressions agree well with the mean free path inferred previously by Faucher-Gigu\`ere et al. (2008a)\nocite{faucher-giguere_evolution_2008}, but have significantly reduced uncertainties. 
On the other hand, this mean free path is larger than that assumed by \cite{madau_radiative_1999} by a factor $\sim2.5$. 
Furthermore, these mean free path measurements are significantly more accurate than the simple model based on the mean spacing between Lyman limit systems assumed by \cite{bolton_observed_2007}. 

\begin{figure*}
\includegraphics[width=1.5\columnwidth]{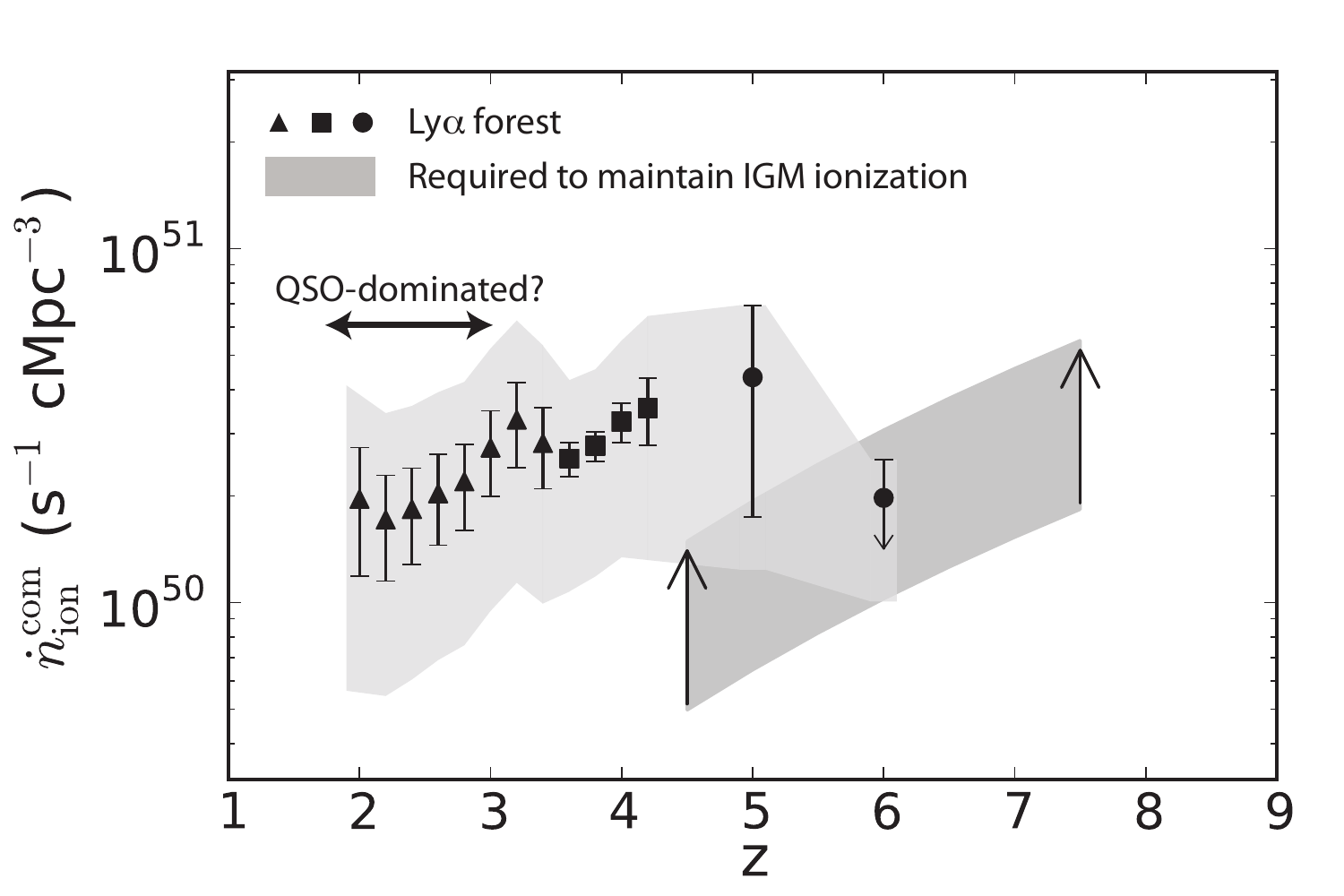}
\caption{Ly$\alpha$ constraints on the rate at which ionizing photons are injected into the IGM (see \S \ref{sec:LyaF} for details and Table \ref{tab:LyaF_data} for numerical values). The light gray band indicates instantaneous constraints from the measured mean transmission of the Ly$\alpha$ forest, including systematic effects. The dark gray band indicates the minimum value necessary to keep the Universe ionized, \textit{assuming that reionization is complete}, for a fiducial IGM temperature $T_{0}=2\times10^{4}$ K and effective clumping factor $C_{\rm HII}=1-3$. Models of the ionizing background indicate that the ionizing emissivity is dominated by star-forming galaxies at $z\gtrsim3$, but that quasars 
may dominate at lower redshifts (Faucher-Gigu\`ere et al. 2008a, 2009).}
\label{fig:LyaF}
\vspace*{0.1in}
\end{figure*}

Figure \ref{fig:LyaF} summarizes the IGM observational constraints on $\dot{n}_{\rm ion}^{\rm com}$. 
The error bars on the data points account for the statistical uncertainty on the photonization rate and on the mean free path. The $z=5$ and $z=6$ error bars also include a systematic error budget on $\Gamma_{\rm HI}$, as quantified by \cite{bolton_observed_2007}.  
Total uncertainties, including systematics, are indicated by the light gray band and estimated as follows. 
First, we allow for a 50\% systematic error on the $\Gamma_{\rm HI}$ data points from Faucher-Gigu\`ere et al. (2008a)\nocite{faucher-giguere_evolution_2008} to account for uncertainties in the thermal state of the IGM and the probability distribution function of density fluctuations, which enter in the mean transmission method (Bolton et al. 2005; Faucher-Gigu\`ere et al. 2008a)\nocite{bolton_lyman_2005, faucher-giguere_evolution_2008}. 
Second, we vary the source spectral index $\gamma$ from 1 to 3. The harder value $\gamma=1$ is preferred by  optical line ratio diagnostics in local starbursts \citep[][]{kewley_theoretical_2001}, while many stellar population synthesis models predict $\gamma\approx3$ \citep[e.g.,][]{leitherer_starburst99:_1999}. 
This range of slopes is also consistent with the possibility that quasars, with mean spectral index $\sim 1.6$ \citep[][]{telfer_rest-frame_2002}, contribute significantly at the lower redshift end. 
The fiducial value assumed in our calculations is $\gamma=1$; since stellar population synthesis models generally predict softer spectra, we do not explore harder values. 
The spectral index of the background in the ionizing regime is not independent but instead satisfies $\gamma_{\rm bg} = \gamma - 3(\beta-1)$,\footnote{This approximation (arising from the frequency dependence of the mean free path; eq. (\ref{local source approximation emissivity})) is valid at least up to $z\sim4$, where the column density distribution has been measured to be well approximated by a series of power laws \citep[][]{prochaska_definitive}. However, it may break down at earlier times, where optically thick absorbers could be relatively more numerous and dominate the opacity. This provides additional motivation for focusing our post-reionization analysis at $z=4$ (\S \ref{sec:evolution}).} where $\beta$ is the slope of the HI column density distribution (Faucher-Gigu\`ere et al. 2008a)\nocite{faucher-giguere_evolution_2008}. We adopt $\beta=1.3$ \citep[][]{songaila_evolution_2010}. 

To indicate the total uncertainty, we first calculate the range of $\dot{n}_{\rm ion}^{\rm com}$ values allowed by simultaneously varying the systematically uncertain parameters to their extremes. We then add the statistical uncertainty to the minimum and maximum values in each redshift bin. 
We believe that this procedure conservatively captures the constraints on $\dot{n}_{\rm ion}^{\rm com}$. 

The mean transmission data points are instantaneous constraints that must be satisfied by the galaxy population. 
Models of the ionizing background at intermediate redshifts indicate that star-forming galaxies dominate at $z \gtrsim 3$ (Haehnelt et al. 2001; Bolton et al. 2005; Faucher-Gigu\`ere et al. 2008a, 2009)\nocite{haehnelt_ionizing_2001, bolton_lyman_2005, faucher-giguere_evolution_2008, faucher-giguere_new_2009}. 
However, quasars may dominate the hydrogen photoionization rate at later times, so that the total background should be regarded as an upper limit to the contribution of star-forming galaxies alone. 
It should also be noted that any redshift evolution in $\dot{n}_{\rm ion}^{\rm com}$ contained in the light gray band in Figure \ref{fig:LyaF} is allowed. In particular, it is possible that the true redshift evolution is titled in slope relative to that suggested by the fiducial data points. 
This is because the uncertain parameters could evolve significantly with redshift. 
For instance, measurements indicate that the IGM temperature peaks at $T_{0} \gtrsim 2\times10^{4}$ K around $z \sim 3.4$ \citep[][]{lidz_measurement_2010}, possibly owing to re-heating from HeII reionization, but could be less than $\sim 10^{4}$ K between HI and HeII reionization \citep[][]{hui_thermal_2003, becker_detection_2011, bolton_improved_2012}. 
Recent observations also indicate that the UV slopes of $z\sim7$ galaxies are significantly bluer than their $z\sim3$ counterparts \citep[][]{bouwens_very_2010}, so that the relevant spectral indexes could also evolve. 

It is apparent from Figure \ref{fig:LyaF} that the best-fit comoving ionizing photon emissivity, $\dot{n}^{\rm com}_{\rm ion}$, increases from $z=2$ to $z=4.2$, and perhaps even to higher redshift. 
This basic behavior stems from the fact that while the photoionization rate, $\Gamma_{\rm HI}$, is approximately constant over this redshift interval, the mean free path decreases rapidly with increasing redshift \citep[][]{faucher-giguere_evolution_2008, mcquinn_lyman_2011}. 
Thus, an increasing ionizing emissivity is required to maintain the observed photoionization rate (eq. \ref{eq:ndotion}). 

\subsection{Keeping the Universe ionized}
\label{sec:keeping_ion}
Figure \ref{fig:LyaF} also shows (as the dark gray band) $\dot{n}_{\rm ion}^{\rm com,crit}$, the minimum $\dot{n}_{\rm ion}^{\rm com}$ required to \textit{keep} the IGM ionized once it has already been reionized. 
This number is obtained by balancing the global recombination rate in the fully ionized IGM with the rate at which ionizing photons escape galaxies: 
\begin{align}
\label{eq:ndot_crit}
\dot{n}_{\rm ion}^{\rm com,crit} & = C_{\rm HII} \alpha_{\rm A}(T_{0}) \bar{n}_{\rm H} ( 1 + Y/4X) (1+z)^{3} \\ \notag
& \approx 3\times10^{50}~{\rm s^{-1}~cMpc^{-3}} \left( \f{C_{\rm HII}}{3} \right) \\ \notag 
& ~~~~~~~~~~~~~~~\times \left(\f{T_{0}}{2 \times 10^4 \, {\rm K}}\right)^{-0.7} \!\! \left( \f{1+z}{7} \right)^{3}.
\end{align}
Here, $\alpha_{\rm A}$ is the case A recombination coefficient of hydrogen. Although we used the case B coefficient for the calculation of the HII volume filling factor during reionization (eq. (\ref{eq:trec})), a large fraction of the recombinations directly to the ground state at later times do not actually contribute to the ionizing background \citep[][]{faucher-giguere_new_2009}.  

The $\dot{n}_{\rm ion}^{\rm com,crit}$ band in Figure \ref{fig:LyaF} covers the range $C_{\rm HI}=1-3$ and assumes a fiducial IGM temperature $T_{0} = 2\times10^{4}$ K. 
This temperature is consistent with measurements at $z \sim 2-4$ \citep[e.g.,][]{lidz_measurement_2010}, but could be higher or lower by $\sim 10^{4}$ K depending on the spectrum of the re-ionizing sources and the time elapsed since reionization \citep[][]{hui_thermal_2003}. 
Estimates of the IGM temperature at $z>4$ in fact suggest $T_{0}<10,000$ K at $z\sim5-6$ \citep[][]{becker_detection_2011, bolton_improved_2012}.
Fortunately, this uncertainty does not substantially affect our arguments, since the recombination coefficient is a relatively weak function of temperature. 

Note that the $\dot{n}_{\rm ion}^{\rm com,crit}$ lower limit only applies after reionization is complete. 
Since we do not \textit{a priori} know the redshift of reionization, the plotted constraint (extending to $z=7.5$) need not necessarily be satisifed everywhere. 
However, any viable reionization scenario must satisfy this constraint at all redshifts following the time when an ionized fraction $\sim 1$ is reached.

\section{Combining Ly$\alpha$ forest, galaxy survey, and WMAP constraints}
\label{sec:evolution}

\begin{figure}
\includegraphics[width=\columnwidth]{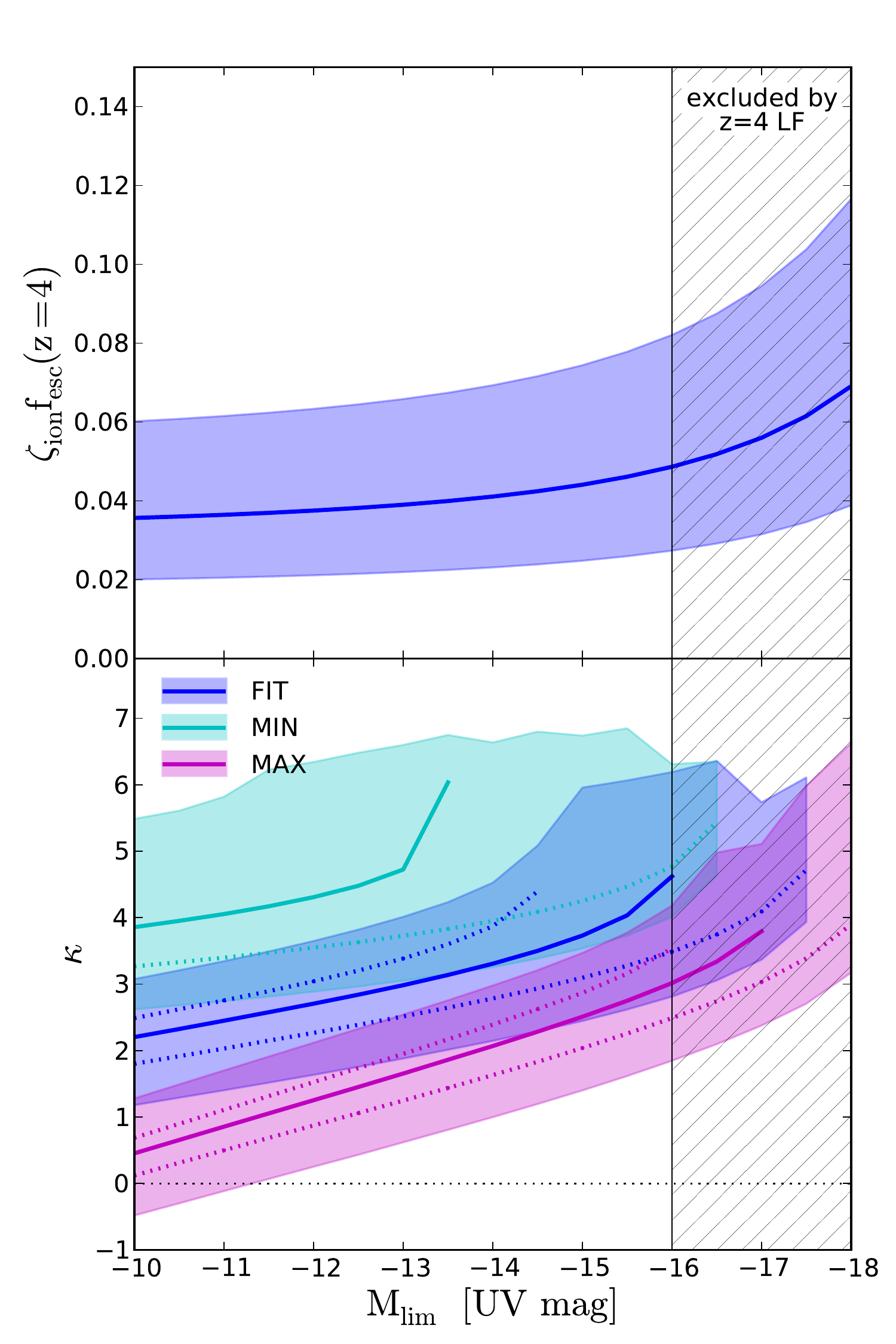}
\caption{
\textit{Top:} Value of $\zetaion \fesc$ at $z=4$ needed to simultaneously match the total comoving emissivity of ionizing photons measured from the Ly$\alpha$ forest, $\ndotion^{\rm com}$, and the observed UV luminosity function at the same redshift, as a function of the limiting UV magnitude. 
Since our LF fits are almost identical at $z=4$, we only show the FIT case. 
Because $\zeta_{\rm ion}=1$ for the fiducial spectral model, the values directly quantify the implied escape fraction. 
\textit{Bottom:} Power-law index $\kappa$ of the redshift evolution of $\zeta_{\rm ion }\fesc$ (see eq.~\ref{eq:power_law}) needed to simultaneously match the $z=4$ \Lya\ forest and \mbox{WMAP-7} Thomson optical depth constraints, as a function of the limiting UV magnitude (assumed constant here), for our three LF evolution fits. 
The solid (dotted) lines correspond to the median value ($\pm1 \sigma$) of the \mbox{WMAP-7} Thomson optical depth. 
The shaded regions encompass the total (including systematic) uncertainty in $\ndotion^{\rm com}(z\!=\!4)$. Note that some models with bright $\Mlim$ do not admit solutions for the entire $\taue$ range. 
}
\label{fig:g0_and_gprime}
\vspace*{0.1in}
\end{figure}

\begin{figure}
\includegraphics[width=\columnwidth]{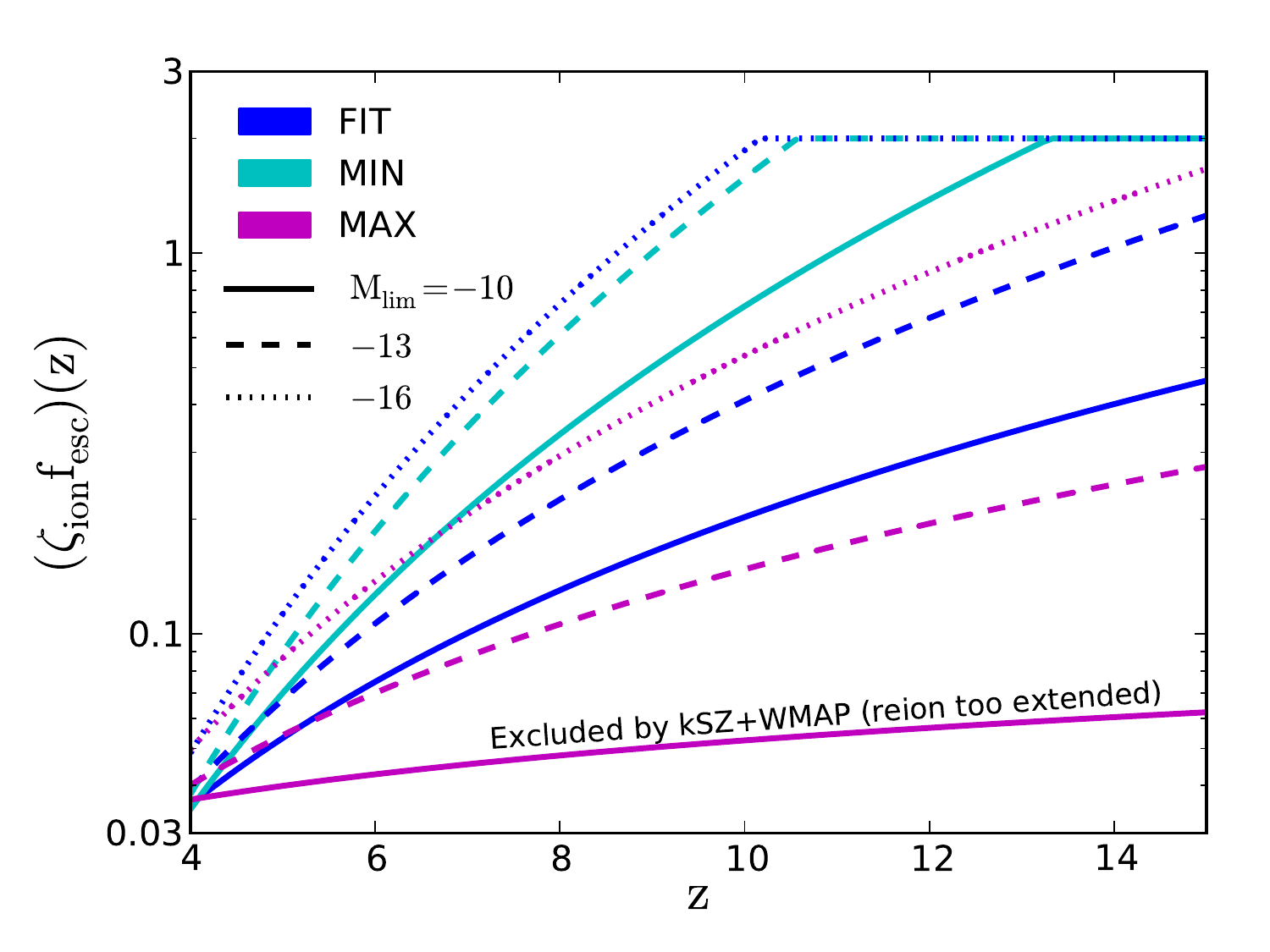}
\caption{Evolution of $\zeta_{\rm ion} \fesc$ versus redshift required to simultaneously satisfy the $z=4$ \Lya\ forest and \mbox{WMAP-7} Thomson optical depth constraints, for $\Mlim = -10, \, -13, \, {\rm and} \, -16$ (models corresponding to the solid lines in Fig.~\ref{fig:g0_and_gprime}). A ceiling of $\zeta_{\rm ion} \fesc < 2$ is imposed in our calculations, corresponding to $f_{\rm esc}=1$ for $\zeta_{\rm ion}=2$ (our HARD spectral model).}
\label{fig:g_vs_z}
\vspace*{0.1in}
\end{figure}

We showed in \S \ref{sec:UF_LV} that many different reionization scenarios are consistent with the existing galaxy survey and WMAP constraints, principally due to uncertainties in $M_{\rm lim}$ and $f_{\rm esc}$, and also $\zeta_{\rm ion}$. We now combine these constraints with the Ly$\alpha$ forest data at lower redshifts (\S \ref{sec:LyaF}), which allow us to break certain degeneracies and quantify possible redshift evolution in the relevant parameters. The key idea is that for any $M_{\rm lim}$, comparison of the Ly$\alpha$ forest and galaxy UV luminosity function data where they overlap imply a unique $\zeta_{\rm ion} f_{\rm esc}$ value (assuming that galaxies dominate the ionizing background). We choose to make this comparison at $z=4$, because at this redshift we expect the hydrogen ionizing background to in fact be dominated by star-forming galaxies (Faucher-Gigu\`ere et al. 2008a, 2009\nocite{faucher-giguere_evolution_2008, faucher-giguere_new_2009}). Furthermore, at this redshift the observational constraints on the Ly$\alpha$ forest transmission and the mean free path of ionizing photons are quite good. Additionally, we do not expect this redshift to be strongly affected by large inhomogeneities in the ionizing background. At higher redshifts (especially at $z\sim6$), interpretation of the Ly$\alpha$ forest data becomes more uncertain because of the small number of sight lines available and because reionization may not be 100\% complete \citep[][]{mcgreer_first_2011}. Because only certain combination $(M_{\rm lim},~\zeta_{\rm ion} f_{\rm esc})$ are allowed at $z=4$, we can quantify whether redshift evolution in these parameters is required in order to simultaneously satisfy the higher-redshift constraints from galaxy surveys and WMAP.

\begin{figure*}
\includegraphics[width=0.75\textwidth]{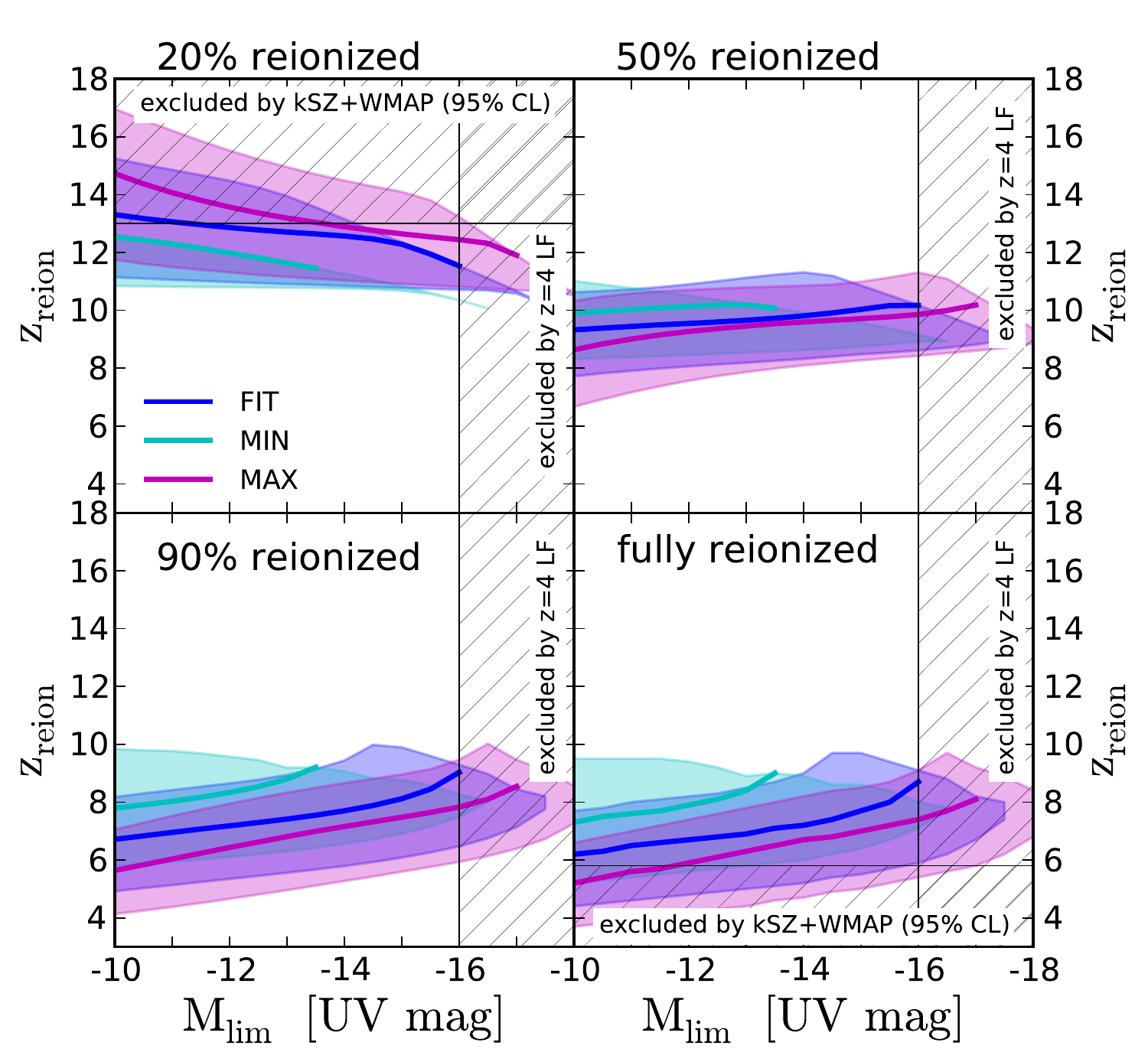}
\caption{Redshift at which reionization is 20\% (top left), 50\% (top right), 90\% (bottom left), and 100\% complete (bottom right), as a function of the limiting UV magnitude, for models in which $\zeta_{\rm ion} \fesc(z)$ is tuned to reproduce both the \mbox{WMAP-7} Thomson optical depth and the $z=4$ \Lya\ forest constraints. The colors represent our three LF evolution fits, and the shaded region encompasses both the \mbox{WMAP-7} $\tau_e$ 1$\sigma$ region and the total (including systematic) uncertainty in $\ndotion^{\rm com}(z=4)$.}
\label{fig:zreion}
\vspace*{0.1in}
\end{figure*}

\begin{figure}
\includegraphics[width=\columnwidth]{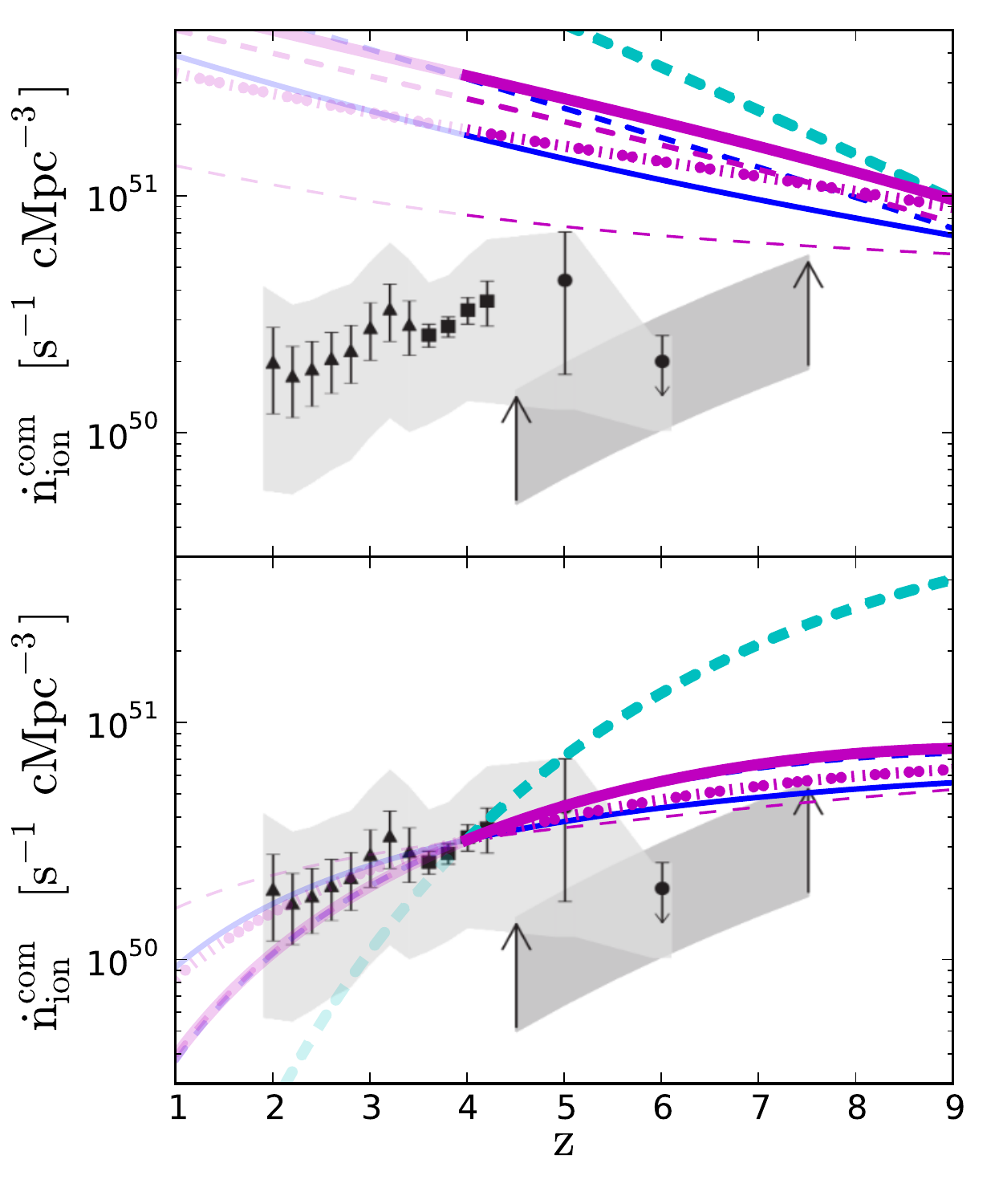}
\caption{Same as Figure~\ref{fig:LyaF}, but with the models from Figure~\ref{fig:QHII_vs_z} overplotted, for the constant $\zeta_{\rm ion} \fesc$ case (top) and modified to allow for redshift evolution in $\zeta_{\rm ion} \fesc$ (bottom). 
For the models with redshift evolution in $\zeta_{\rm ion} \fesc$, the luminosity function parameterization and $M_{\rm lim}$ are fixed to the values from Figure 4, $\zeta_{\rm ion} \fesc(z=4)$ is set by the Ly$\alpha$ forest data at $z=4$, and we solve for $\kappa$ (eq. (\ref{eq:power_law_kappa})) such that the Thomson optical depth matches the central WMAP-7 measurement. 
In the same order as the legend in Figure~\ref{fig:QHII_vs_z}, the best-fit values are $\kappa = 2.2, \, 3.3, \,  3.0, \,  1.3, \, 6.7, \, 2.1, \, 3.0$. The model lines are greyed-out at $z<4$, since we do not utilize galaxy luminosity function data at those redshifts.}
\label{fig:LyaF_with_models}
\vspace*{0.1in}
\end{figure}

In principle, $M_{\rm lim}$, $\zeta_{\rm ion}$, and $f_{\rm esc}$ can all be arbitrary functions of redshift.  However, the limited data do not allow us to explore the entire range of possibilities. Instead, we adopt simple, single-parameter power-law models to quantify the required redshift evolution:
\begin{eqnarray}
\label{eq:power_law_kappa}
(\zetaion \fesc)(z) \!\!\! & = & \!\!\! (\zetaion \fesc)(z\!=\!4) \, \left( \f{1+z}{5} \right)^{\kappa} \\
10^{-0.4 \Mlim(z)} \!\!\! & = & \!\!\! 10^{-0.4 \Mlim(z\!=\!4)} \, \left( \f{1+z}{5} \right)^{- \lambda} . 
\label{eq:power_law}
\end{eqnarray}
The last expression encodes a power-law parameterization in the UV luminosity, which is logarithmically related to the UV magnitude. 
In order to prevent unphysical values of $\fesc$, we impose $\zeta_{\rm ion} \fesc \leq 2$, corresponding to a ceiling of $\fesc=1$ for the case $\zetaion=2$ (our HARD spectral model; \S \ref{sec:zeta_ion}). 
There is also a physical lower limit on the luminosity of the faintest galaxies, corresponding to the minimum halo mass in which baryons can collapse and form stars. 
However, this limit is not as well understood and could depend significantly on redshift. 
We impose an extremely faint ceiling of $\Mlim=0$ ($\sim 4$ orders of magnitude below the UV suppression sacle of $M = -10$), and discuss the viability of different scenarios later. 
Note that we have defined the signs of the $\kappa$ and $\lambda$ power-law indexes so that positive values correspond to increasing efficiency of ionizing photon production going to higher redshifts. 

In the following, we consider possible evolution first in $\zetaion\fesc$ (\S \ref{sec:g_evo}), and then in $\Mlim$ (\S \ref{sec:Mlim_evo}). In principle, there could be simultaneous evolution in both $\zetaion\fesc$ and $\Mlim$, but the data do not allow us to discriminate between such mixed scenarios. 
Furthermore, we will show that the strong redshift evolution required in models relying only on relatively bright galaxies is most plausibly accounted for by evolution in the escape fraction. 
In \S \ref{sec:reionization_duration}, we consider the redshift and duration of reionization in different allowed scenarios, showing that the redshift of 50\% ionized fraction is consistently at $z_{\rm reion}(50\%) \sim 10$ among the different models allowed by the data, but that the duration of reionization, $\Delta z_{\rm reion} \equiv z_{\rm reion}(100\%) - z_{\rm reion}(20\%)$ (where $z_{\rm reion}(x)$ is the redshift such that $Q_{\rm HII}(z_{\rm reion}(x))=x$), is on the other hand sensitive to the contribution of faint galaxies.

\subsection{Redshift evolution of $\zeta_{\rm ion} \fesc$}
\label{sec:g_evo}
We first focus on the redshift evolution of $\zeta_{\rm ion} f_{\rm esc}$ from $z=4$ toward higher redshifts.

Our goal is to determine as a function of $\Mlim$ (here assumed to be independent of $z$) what values of $\kappa$ are consistent with both the $z\approx 4$ \Lya\  forest and \mbox{WMAP-7} Thomson optical depth constraints, 
while also satisfying the measurements of the galaxy UV LF at the bright end. Operationally, we first determine for a given $\Mlim$ what range of $\zeta_{\rm ion} \fesc(z=4)$ is required to give $\ndotion^{\rm com}(z=4) = 3.2^{+2.2}_{-1.4} \times 10^{50} \, {\rm s}^{-1} \, {\rm cMpc}^{-3}$. The results of this calculation are shown in the top panel of Figure~\ref{fig:g0_and_gprime}. For $\Mlim=-10$ to $-16$, the \Lya\ constraints require $\zeta_{\rm ion} \fesc(z=4)$ to lie between 2\% and 8\% \citep[since the luminosity function has already been measured down to $M_{\rm UV}=-16$ at $z=4$, cases with brighter $M_{\rm lim}$ at this redshift are not allowed;][]{bouwens_uv_2007}. As $\zeta_{\rm ion}=1$ for the fiducial spectral model, this directly quantifies the implied escape fraction. This result is independent of which LF fit we employ, since at $z=4$ the parameters of our three fits are nearly identical.

In the second step, we determine for each $\Mlim$ and $\zeta_{\rm ion} \fesc(z=4)$ what values of the power-law index $\kappa$ yield a Thomson optical depth in the range allowed by \mbox{WMAP-7}. The resulting range of allowed $\kappa$ values is shown in the bottom panel of Figure~\ref{fig:g0_and_gprime}, with the different color bands corresponding to our three LF fits. The width of the bands encompasses both the 1$\sigma$ uncertainty of the \mbox{WMAP-7} Thomson optical depth measurement and the total (including systematic) uncertainty in the $z=4$ \Lya\ forest data. 
Figure \ref{fig:g_vs_z} shows the curves of $\zeta_{\rm ion} f_{\rm esc}$ versus $z$ corresponding to allowed values of $\kappa$, for representative choices of $M_{\rm lim}$. For each $\kappa$ solution, Figure \ref{fig:zreion} shows $z_{\rm reion}(20\%)$, $z_{\rm reion}(50\%)$, $z_{\rm reion}(90\%)$, and $z_{\rm reion}(100\%)$.

Lastly, Figure \ref{fig:LyaF_with_models} shows explicit examples of how redshift evolution in $\zeta_{\rm ion} f_{\rm esc}$ allows models to simultaneously satisfy reionization-epoch constraints and the $z<6$ Ly$\alpha$ forest data. 
Specifically, we consider the same models as in Figure~\ref{fig:QHII_vs_z} (top) and modify them to allow for redshift evolution in $\zeta_{\rm ion} \fesc$ (bottom). 
For the models with redshift evolution in $\zeta_{\rm ion} \fesc$, the luminosity function parameterization and $M_{\rm lim}$ are fixed to the values from Figure 4, $\zeta_{\rm ion} \fesc(z=4)$ is set by the Ly$\alpha$ forest data at $z=4$, and we solve for $\kappa$ such that the Thomson optical depth matches the central WMAP-7 measurement. 
While the original models without evolution in $\zeta_{\rm ion} \fesc$ overproduce the ionizing emissivity probed by the Ly$\alpha$ forest at $z=4$, the modified models simultaneously satisfy all the constraints. 
Furthermore, models exist in which the extrapolation to $z=2$ is also in good agreement with the lower-redshift Ly$\alpha$ forest data.

At the $1\sigma$ level, only the MAX model with very faint $\Mlim \gtrsim -11$ can accommodate no redshift evolution in $\zeta_{\rm ion} f_{\rm esc}$. However, these scenarios are disfavored by external constraints on the duration of reionization from the kinetic Sunyaev-Zeldovich effect (Zahn et al. 2011), which in combination with the WMAP-7 optical depth constrains the timing of its beginning and end. 
For the most conservative case of arbitrary correlations between the thermal Sunyaev-Zeldovich (tSZ) effect and the cosmic infrared background (CIB), \cite{zahn_comic_2011} find that $z_{\rm reion}(20\%)<13.1$ at 95\% confidence level (CL), and $z_{\rm reion}(99\%)>5.8$, also at 95\% CL. 
In this work, we take these constraints at face value. 
It is important to bear in mind, however, that the templates on which they are based assume that reionization occurs primarily via star-forming galaxies. 
Furthermore, the limits on the kSZ signal rely critically on accurate subtraction of contaminating point sources. 
It will thus be important to confirm these findings with refined analyses.
 
For the best-fit parameterization of the UV LF (the FIT model), models with no redshift evolution in $\zeta_{\rm ion} f_{\rm esc}$ are disfavored even for $M_{\rm lim}=-10$. 
Models that rely only on brighter galaxies formally satisfy all the present constraints but only for strong redshift evolution in $\zeta_{\rm ion} f_{\rm esc}$. 
For example, the case of $M_{\rm lim}=-16$ for the FIT parameterization requires an evolution in $\zeta_{\rm ion} f_{\rm esc}$ by a factor $\approx20$ from $z=4$ to $z=9$. 
As we will discuss at greater length in \S \ref{sec:conclusions}, models that rely too heavily on fainter galaxies may be in tension with theoretical models that suppress star formation in early, low-mass systems \citep[e.g.,][]{krumholz_metallicity-dependent_2011, kuhlen_dwarf_2012}, which are helpful in explaining some properties of the cosmic star formation history. 
If star formation is indeed suppressed in those early dwarfs, then the existing data would imply strong evolution in the escape fraction. 

We also explored constraints on the redshift evolution $\zeta_{\rm ion} f_{\rm esc}$ from $z=2$ to $z=4$ by comparing the Ly$\alpha$ forest data to the galaxy UV luminosity function from \cite{reddy_steep_2009} at $z=2$. 
Over that redshift interval, a wide range $\kappa \sim 0-4.5$ is allowed, almost independent of the assumed $M_{\rm lim}$ owing to the relatively shallow faint-end slope of the luminosity function. 
In particular, the combination of the UV luminosity function and Ly$\alpha$ forest data alone do not require any significant evolution. 
Note that such evolution is nonetheless allowed by the data, and in fact suggested by direct Lyman continuum observations \citep[e.g.,][]{inoue_escape_2006, siana_deep_2010}. 
Although this is not necessary on physical grounds, it is interesting that most of the $\kappa$ values implied from $z=4$ and up (Fig. \ref{fig:g0_and_gprime}) are also allowed from $z=2$ to $z=4$.  

\subsection{Redshift evolution of $\Mlim$}
\label{sec:Mlim_evo}

\begin{figure}
\includegraphics[width=\columnwidth]{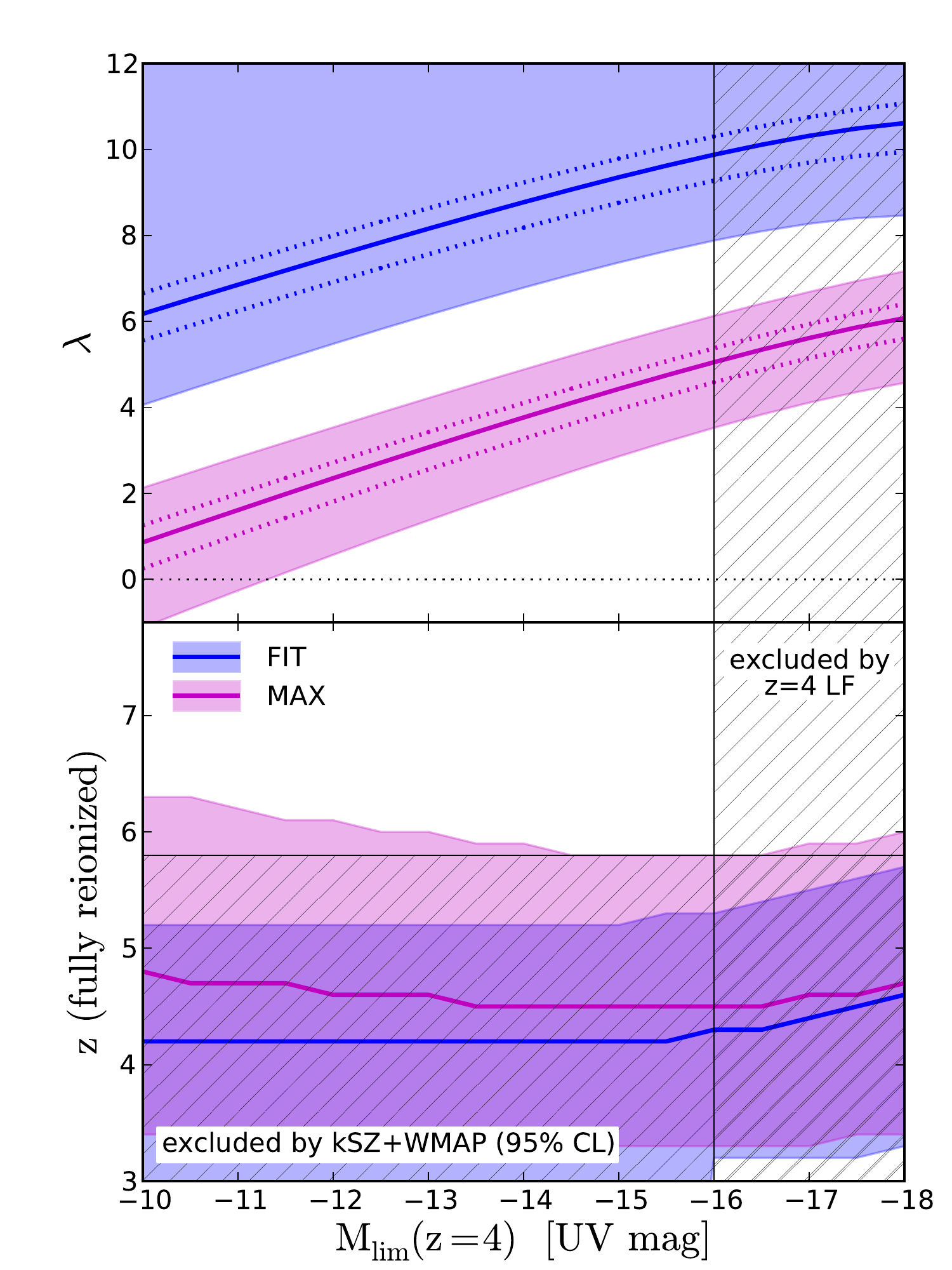}
\includegraphics[width=\columnwidth]{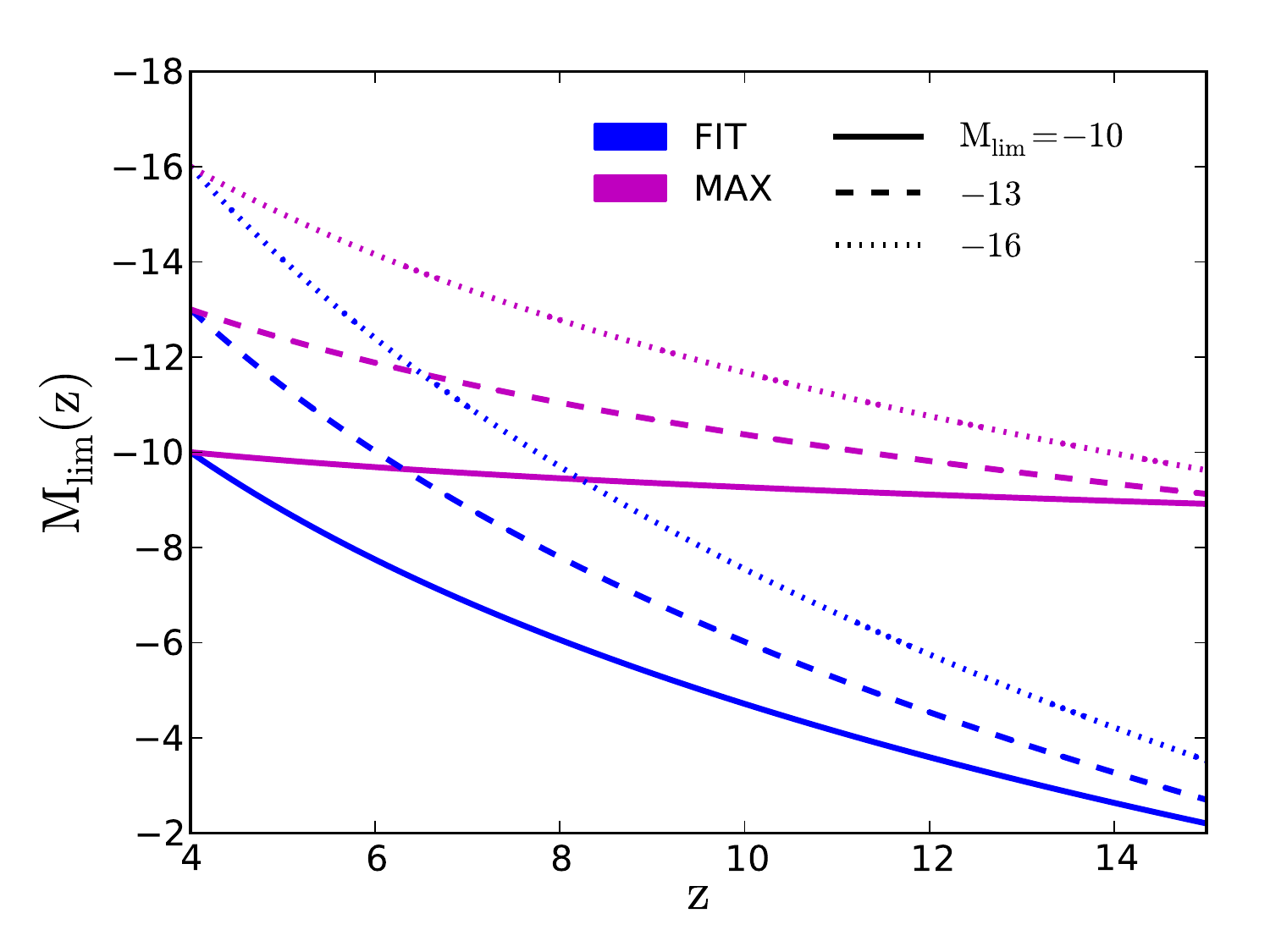}
\caption{\textit{Top:} Power-law index $\lambda$ of the redshift evolution of the limiting UV magnitude $\Mlim$ (see eq. 14)  needed to simultaneously match the $z=4$ \Lya\ forest and the \mbox{WMAP-7} Thomson optical depth constraints, as a function of $\Mlim(z=4)$ ($\zetaion \, \fesc$ is assumed constant). As before, the solid (dotted) lines correspond to the median value ($\pm1\sigma$) of the \mbox{WMAP-7} Thomson optical depth, 
and the shaded regions encompass the total (including systematic) uncertainty in $\ndotion^{\rm com}(z\!=\!4)$. Models with the MIN LF fit do not admit any solutions with evolution only in $\Mlim$. 
\textit{Middle:} Redshift at which reionization is completed ($\QHII=1$). 
Most of the $\Mlim$-only evolution models do not complete reionization by $z=5.8$, as required by the combination of kSZ and WMAP-7 data \citep[][]{zahn_comic_2011}. \textit{Bottom:} Median $\Mlim$ redshift evolution (corresponding to the solid lines in the top panel), for $\Mlim = -10, \, -13, \, {\rm and} \, -16$. Very faint limiting magnitudes $M_{\rm lim} \gg -10$ are also likely excluded on physical grounds.}
\label{fig:dMlimdz}
\end{figure}

We now turn to the possibility of redshift evolution in $\Mlim$. Starting from the same values of $\zeta_{\rm ion} \fesc(z=4)$ for a given $\Mlim(z=4)$ as in the previous section, we determine what values of $\lambda$ can produce agreement between the \Lya\ forest constraints at $z=4$, and higher-redshift LF and WMAP-7 constraints. The top panel of Figure~\ref{fig:dMlimdz} demonstrates that substantial evolution is necessary to match the \mbox{WMAP-7} Thomson optical depth measurement. Many models actually do not allow for a solution: for the MIN LF fit, the contribution of dwarfs is suppressed to such a degree that no amount of $\Mlim$ evolution is able to raise $\taue$ into the range allowed by \mbox{WMAP-7}. The FIT model only has solutions with very steep $\Mlim$ evolution, requiring $\Mlim$ significantly below $-10$ at high redshift. Only the MAX models are able to provide solutions with a moderate amount of $\Mlim$ evolution. 
Most importantly, as the middle panel of Figure~\ref{fig:dMlimdz} demonstrates, the majority of the models with evolving $M_{\rm lim}$ (but constant $\zeta_{\rm ion} f_{\rm esc}$) do not reach complete reionization by $z=5.8$, as required by the combination of the kSZ and WMAP constraints \citep[][]{zahn_comic_2011}. 

Except for small, extreme corners of parameter space, it is therefore not possible to simultaneously match $z=4$ \Lya\ forest and reionization constraints, and complete reionization in time, by allowing only evolution in $\Mlim$. 
This provides further evidence for the need for a significant redshift evolution in $\zeta_{\rm ion} f_{\rm esc}$ from $z=4$ toward higher redshifts, as discussed in the previous section. 
In contrast, the data do not provide conclusive evidence for a significant redshift evolution in $M_{\rm lim}$ (although some evolution may be expected on physical grounds). 
We will therefore concentrate the following discussion on scenarios with constant $M_{\rm lim}$ but evolving $\zeta_{\rm ion} f_{\rm esc}$.  
It is important to keep in mind, however, that evolution in $\Mlim$ could reduce the required amount of evolution in $\zeta_{\rm ion} \fesc$ somewhat. 

\subsection{Redshift and duration of reionization}
\label{sec:reionization_duration}
Interestingly, Figure \ref{fig:zreion} shows that the redshift of 50\% ionized fraction is consistently at $z\sim10$ between the different allowed scenarios. 
This is consistent with the redshift of instantaneous reionization implied by WMAP-7, $z_{\rm reion}=10.4\pm1.2$. 
However, the predicted duration of the reionization process is more extended for scenarios that include a larger contribution from faint galaxies. 
This is a consequence of the shape of the LF: the brighter and closer to $M^*$ (the knee of the LF) $\Mlim$ is, the fewer galaxies contribute to reionization at high $z$. The reionization process then quickly sets in once the exponential cutoff of the LF has shifted to bright enough galaxies that $\Mlim$ galaxies become common. In contrast, with a faint $\Mlim$ abundant faint galaxies contribute to reionization even at very high redshifts, and therefore the overall evolution of the process is slowed down. 

As shown in Figure \ref{fig:zreion}, the kSZ data favor relatively short reionization histories and thus disfavor models that rely too heavily on high-redshift dwarfs, as in the MAX models.  
A range of MIN and FIT models are however allowed. 
In particular, our conclusions are consistent with theoretical analyses anchored in the predicted dark matter halo mass function, rather than to the observed luminosity function \citep[][]{trenti_galaxy_2010,mesinger_kinetic_2011, ciardi_effect_2012}, indicating that as of yet undetected galaxies must contribute significantly to reionization. 
However, our analysis suggest alternative possibilities when strong redshift evolution in $\zeta_{\rm ion} f_{\rm esc}$ is allowed.

Recently, two significant observational advances in using astrophysical sources to probe the epoch of reionization have been reported. First, \cite{mortlock_luminous_2011} discovered a luminous quasar at $z=7.085$ in which the Ly$\alpha$ transmission profile is consistent with an IGM neutral fraction $\sim10$\% at that time. Although this is compatible with most of the scenarios allowed by our analysis (Fig. \ref{fig:zreion}), excluding those relying on a maximal contribution from dwarf galaxies, alternative interpretations exist in the context of inhomogeneous reionization \citep[][]{bolton_neutral_2011}. 

Second, recent surveys for Ly$\alpha$ emitting galaxies at $z \geq 6$ have found a decreasing fraction of Lyman break-selected galaxies with detected Ly$\alpha$ emission of rest-frame equivalent width $\geq 20$~\AA~from $z\sim6$ to $z\sim7$ \citep[][]{schenker_keck_2011, pentericci_spectroscopic_2012, ono_spectroscopic_2012}. 
The decline is such that existing models of Ly$\alpha$ propagation through galactic winds and the intervening IGM indicate $Q_{\rm HII}\gtrsim50$\% at $z\sim7$ \citep[][]{dijkstra_detectability_2011}. Such a high neutral fraction at $z=7$ is in tension with the concordance scenarios summarized in Figure \ref{fig:zreion}. Furthermore, if reionization is essentially complete by $z\sim6$ as the combination of kSZ and WMAP-7 measurements indicate, and as suggested by Gunn-Peterson troughs in the Ly$\alpha$ forest \citep[e.g.,][]{fan_evolution_2002}, then a very rapid evolution in the neutral fraction would be implied. These findings are not easy to reconcile, and highlight the need for more detailed modeling of Ly$\alpha$ radiative transfer in order to fully exploit the new and upcoming high-quality data.

\section{Summary and Discussion}
\label{sec:conclusions}
Measurements of the galaxy UV ($\sim$1,500~\AA) luminosity function at redshifts $z \gtrsim 6$, recently improved by more than an order of magnitude thanks to the WFC-3 camera on HST, provide constraints on the likely sources of hydrogen reionization. 
However, these observations only directly reveal the sources luminous enough to be individually detected. 
Furthermore, converting the luminosity function measurements to IGM ionization rates involves large uncertainties, in addition to the extrapolation necessary to model sources too faint to be detected, including the SED of the star-forming galaxies and their escape fraction of ionizing photons. 
These latter uncertainties are encapsulated in the dimensionless factor $\zeta_{\rm ion} f_{\rm esc}$ used to convert from a luminosity at 1,500~\AA~to a rate of production of ionizing photons escaping into the IGM.

Given these uncertain parameters, we showed in \S \ref{sec:UF_LV} that many scenarios exist in which star-forming galaxies are the dominant ionizing sources and which satisfy both the galaxy survey constraints and the Thomson optical depth implied by the \mbox{WMAP-7} data. Such scenarios include ones with escape fraction ranging from $f_{\rm esc}=5$\% to $f_{\rm esc}=50$\%, and limiting UV magnitude ranging from $M_{\rm lim}=-16$ to $M_{\rm lim}=-10$, even when these are assumed to be constant. 
Thus, these constraints alone \citep[which have been the focus of many analyses; e.g.,][]{bouwens_lower-luminosity_2011, bunker_contribution_2010} are not sufficient to determine the role of faint galaxies in reionizing the Universe, and whether such galaxies are even present in significant number. 

In \S \ref{sec:LyaF} we used the Ly$\alpha$ forest at redshifts $2 \leq z \leq 6$ to measure the total instantaneous rate at which ionizing photons are injected into the IGM. Although these measurements cover redshifts past the epoch of reionization, they provide significant leverage over galaxy surveys. In particular, the total ionizing emissivity implied by the mean transmission of the Ly$\alpha$ forest does not rely on assuming an escape fraction or a limiting magnitude, two of the main uncertainties limiting the predictive power of UV luminosity function measurements alone. At $z=4$, where the Ly$\alpha$ forest data is both abundant and free of large systematic effects due to inhomogeneities in the ionizing background, comparison with the UV luminosity function allowed us to determine 
$\zetaion \fesc$ almost independently of $\Mlim$, owing to the comparatively shallow faint-end slope of the LF at $z \la 4$. Since $M_{\rm lim} \geq -16$ at $z=4$ (the UV luminosity function having already been measured down to this magnitude), $f_{\rm esc}(z=4) = 2-8\%$ (median $f_{\rm esc}(z=4) = 4\%$), for the fiducial spectral model $\zeta_{\rm ion}=1$.

Combining the Ly$\alpha$ forest, \mbox{WMAP-7}, and galaxy survey data and assuming that galaxies are the main ionizing sources requires either: 1) extrapolation of the galaxy luminosity function down to very faint UV magnitudes $M_{\rm lim}\sim-10$, corresponding roughly to the UV background suppression scale (e.g., Faucher-Gigu\`ere et al. 2011; but see Dijkstra et al. 2004\nocite{faucher-giguere_baryonic_2011, dijkstra_photoionization_2004}); 2) an increase of the escape fraction by a factor $\gtrsim10$ from $z=4$ to $z=9$; or 3) more likely, a hybrid solution in which undetected galaxies contribute significantly and the escape fraction increases more modestly. 

The present data do not allow us to select a unique viable reionization scenario. 
Quantitatively, a range of combinations of limiting magnitudes and redshift evolution of the parameters affecting the conversion of $\sim1,500$~\AA~UV luminosity functions to rates of production of ionizing photons are allowed and summarized in Figures \ref{fig:g0_and_gprime}, \ref{fig:g_vs_z}, and \ref{fig:dMlimdz}. 
Redshift evolution in the limiting magnitude alone requires extreme assumptions in order to satisfy both the Ly$\alpha$ forest and WMAP-7 constraints without appealing to very faint galaxies. 
Even so, such scenarios predict that reionization ends at $z\lesssim 6$, in tension with recent measurements of the kinetic Sunyaev-Zeldovich effect by SPT, which indicate that reionization ends earlier than $z=5.8$ at 95\% CL \citep[][]{zahn_comic_2011}. 
On the other hand, significant redshift evolution in $\zeta_{\rm ion} f_{\rm esc}$ is more plausible. 
In fact, $\zeta_{\rm ion}$ can vary by a factor $\sim 4$ owing to changes in the age, metallicity, and IMF of the stellar populations (corresponding to the range $\zeta_{\rm ion}=0.5-2$ assumed in this paper; \S \ref{sec:zeta_ion}).\footnote{Consistency between star formation rate and stellar mass density measurements at $z\sim7-8$ suggests that the IMF of reionization-epoch galaxies is not very different from the local Universe \citep[][]{bouwens_ultraviolet_2011}.} More importantly, $f_{\rm esc}$ can in principle increase from $f_{\rm esc} \sim 4$\% at $z=4$ to $f_{\rm esc} \sim 1$ at earlier times. 
A similar strong redshift evolution of the ionizing luminosity-weighted escape fraction was also recently found to be required in the ``minimal cosmic reionization model'' of \citet{haardt_radiative_2011}.

Although there are at present no direct constraints on the escape fraction from faint galaxies during the epoch of reionization, deep searches for escaping Lyman continuum radiation at lower redshifts do show some evidence for redshift evolution \citep[][]{steidel_lyman_2001, inoue_escape_2006, shapley_direct_2006, cowie_measuring_2009, siana_deep_2010, nestor_narrowband_2011}. 
Such evolution, in which the escape fraction increases with redshift, could owe to increased feedback at earlier times when star formation was more vigorous \citep[e.g.,][]{wise_ionizing_2009}. 
This picture, in which ionizing photons escape galaxies along lines of sight cleared of obscuring gas, would be consistent with Lyman continuum observations suggesting ``on/off'' escape, possibly connected to the viewing geometry \citep[e.g.,][]{shapley_direct_2006, nestor_narrowband_2011, vanzella_detection_2012}. 
Another possibility is that faint galaxies may typically have higher escape fraction than more massive galaxies \citep[e.g.][]{yajima_escape_2011}, in which case the larger relative abundance of faint galaxies at high redshift would result in an increase in the population-averaged escape fraction.
The extremely blue UV continuum slopes recently reported for $z\sim7$ galaxies \citep[][]{bouwens_very_2010} are also suggestive of weak nebular recombination emission, which would be consistent with very high escape fractions of ionizing photons (but see Dunlop et al. 2011 for a critical analysis of the UV continuum slopes)\nocite{dunlop_critical_2011}. 

Recent models predict that star formation is suppressed in low-mass, high-redshift galaxies owing to the metallicity dependence of the transition from warm HI to dense molecular gas. In the fiducial implementations of \cite{krumholz_metallicity-dependent_2011} and \cite{kuhlen_dwarf_2012}, metallicity effects can strongly suppress star formation in reionization-epoch galaxies in haloes of mass of $M_{\rm h}\sim 10^9 - 10^{10}$ M$_{\odot}$, corresponding to $M_{\rm UV}\sim -13$ to $-16$ at $z\sim7$. 
Formally, even the brighter $\Mlim$ of these models can satisfy the existing reionization constraints, but only for strong redshift evolution in $\zeta_{\rm ion} f_{\rm esc}$ (Fig. \ref{fig:g0_and_gprime}). 
Since such a H$_{2}$-regulated star formation suppression threshold is not far from the current limits of HST observations at $z\sim7$, deeper integrations have the potential to significantly constrain those models. 
As \cite{kuhlen_dwarf_2012} showed, the exact halo mass below which metal-poor dwarfs are suppressed is however sensitive to the details of the model implementation. 
In particular, the relevant mass scale depends significantly on the metallicity floor assumed to model the unresolved effects of metal enrichment by early Pop III stars. 
If the correct halo mass threshold is lower by an order of magnitude relative to the fiducial models of \cite{kuhlen_dwarf_2012} and closer to the predictions of \cite{krumholz_metallicity-dependent_2011}, i.e. $M_{\rm h} \sim 10^{9}$ M$_{\odot}$ at $z\sim7$, then abundance matching suggests that the turn over in the luminosity function would occur instead at $M_{\rm UV} \sim -13$ (see Fig.~\ref{fig:abundance_match}). 
For limiting magnitudes in this neighborhood, more modest redshift evolution in $\zeta_{\rm ion} f_{\rm esc}$ can satisfy the galaxy survey, WMAP-7, and Ly$\alpha$ forest data. 
Thus, given the present implementation uncertainties, H$_{2}$-regulated star formation in high-redshift dwarfs is consistent with galaxies reionizing the Universe. 

For each scenario satisfying the observational constraints considered in this work, we evaluated the timing and duration of the corresponding reionization history (Fig. \ref{fig:zreion}). 
Interestingly, the redshift at which the ionized fraction reaches 50\% is consistent among the different allowed scenarios, $z_{\rm reion}(50\%) \sim 10$. 
This redshift is consistent with the redshift of instantaneous reionization $z_{\rm reion}=10.4\pm1.2$ implied by the \mbox{WMAP-7} analysis \citep[][]{komatsu_seven-year_2011}. 
On the other hand, the predicted duration of reionization is quite sensitive to the fractional contribution of faint galaxies. 
For instance, our MAX parameterization with $M_{\rm lim}=-10$ (i.e. with a heavy dwarf contribution) allows scenarios with $z_{\rm reion}(20\%)=17$ and $z_{\rm reion}(100\%)<4$. 
On the other hand, models which rely on bright galaxies are predicted to reionize the Universe much more sharply, with $z_{\rm reion}(20\%)-z_{\rm reion}(100\%) \approx 3$ for our FIT parameterization with ($M_{\rm lim}=-16$). 
Thus, experiments capable of measuring the duration of reionization will have direct implications for faint galaxies. 
In fact, the more extended reionization histories which rely on a heavy contribution from faint galaxies are already ruled out by recent constraints from the kinetic Sunyaev-Zeldovich effect by SPT. 
In the next few years, expanded data sets from high-resolution microwave background experiments in combination with more precise measurements of the integrated Thomson optical depth with \emph{Planck}\footnote{http://www.rssd.esa.int} will improve these constraints further. 
Refined analyses should also improve the accuracy with which contaminating point sources are subtracted, and thus solidify the results. 

On the theoretical front, there is much room for improving our understanding of how to reliably use astrophysical sources such as high-redshift Ly$\alpha$ emitting galaxies, luminous quasars, and $\gamma-$ray bursts to measure the neutral fraction in the IGM.
Finally, efforts aimed at detecting 21 cm emission from high-redshift intergalactic neutral gas \citep[e.g.,][]{bowman_field_2007, parsons_precision_2010, bowman_lower_2010} are poised to eventually directly map the reionization epoch.

\section*{Acknowledgments}
CAFG is supported by a fellowship from the Miller Institute for Basic Research in Science and NASA grant 10-ATP10-0187. 

\bibliographystyle{mn2e}
\bibliography{DwarfsReionization}

\end{document}